\newcommand{\beq}{\begin{equation}}
\newcommand{\eeq}{\end{equation}}

\newcommand{\vet}[1]{\textbf{#1}}
\newcommand{\surv}[2]{\mathbbm 1_{\{\tau_{#1}>#2\}}}
\newcommand{\expV}[2]{\mathbb E_{#2} \left[{#1} \right]}
\newcommand{\survInd}[1]{\mathbbm 1_{\{\tau>#1\}}}

\documentclass[12pt]{article}
\usepackage[english]{babel}
\usepackage{graphicx}
\usepackage{graphics}
\usepackage{epsfig}
\usepackage{epstopdf}
\usepackage{times}
\usepackage{a4wide}
\usepackage{amsmath}
\usepackage{amsfonts}
\usepackage{amssymb} 
\usepackage{commath}
\usepackage{amsthm}
\usepackage{amssymb,bbm}
\usepackage{enumerate}
\usepackage[round]{natbib}
\usepackage{subfig}
\usepackage{slashbox} 
\usepackage{floatrow}
\newfloatcommand{capbtabbox}{table}[][\FBwidth]

\usepackage{blindtext}

\newtheorem{myDef}{Definition}

\newtheorem*{prop}{Proposition}
\DeclareMathOperator{\corr}{corr}               
\DeclareMathOperator{\sign}{sign}               
\DeclareMathOperator{\std}{std}               
\DeclareMathOperator{\var}{var}               

\author{Damiano Brigo\thanks{Imperial College, London, U.K.
({\texttt{damiano.brigo@imperial.ac.uk}})} \and Jo{\~{a}}o~Garcia\thanks{The~opinions~expressed~in this paper are those of the authors and do not
necessarily reflect those of their employers. The authors would also like to thank and acknowledge the comments of Thomas Aubrey (formerly Managing Director in Fitch Solutions) that improved significantly the quality of this work. Any possible remaining errors are those of the authors.}
\and\
Nicola Pede\thanks{Imperial College, London, U.K.
({\texttt{n.pede@imperial.ac.uk}})}
}

\title{
CoCo Bonds Valuation with Equity- and Credit-Calibrated First Passage Structural Models
}

\date{First version: Jan 31st, 2011\\
Second version: May 4th, 2012. This version: \today}

\begin{document}
\maketitle
\begin{abstract}
After the beginning of the credit and liquidity crisis,  financial institutions have been considering creating a convertible-bond type contract focusing on Capital. Under the terms of this contract, a bond is converted into equity if the authorities deem the institution to be under-capitalized. This paper discusses this {\emph{Contingent Capital} (or Coco) bond} instrument and presents a pricing methodology based on firm value models. The model is calibrated to readily available market data. A stress test of model parameters is illustrated to account for potential model risk. Finally, a brief overview of how the instrument performs is presented.   
   
\end{abstract}

\medskip \noindent { AMS Classification Code: 91B70},\ \ \  {JEL Classification Code: G13}


\bigskip

\pagestyle{myheadings} \markboth{}{{\footnotesize  D. Brigo, J. Garcia and N. Pede. CoCo bonds valuation with firm value models}}

\section{Introduction}
\label{sec:Introduction}

The credit crisis that began in Summer 2007 has led the financial industry to reconsider deeply ingrained concepts.  One of those ideas is the role of capital and leverage in financial institutions. The term \emph{capital} is related to the amount of assets and cash an institution is supposed to set aside to prevent its net asset value from falling below the level that could affect its business or hinder the pursuit of its strategic objectives. Capital has been the focus of regulation. There are broadly two notions of capital: \emph{regulatory} (RC) and \emph{economic} (ECAP) Capital. The former is the capital mandated by the regulatory authorities for the financial institution to be considered safe. The latter is an institution's internal estimate of the needed capital and is supposed to reflect more closely the market and economic conditions in which the company operates. 

\subsubsection*{Regulatory Capital assumptions: Granularity and Additivity}
Up until the credit crisis, the regulatory capital requirement for a position in the portfolio of a financial institution was based on two key technical assumptions (see ~\cite{Gordy2003}): \emph{portfolio invariance} and \emph{additivity}. The first means that the cost of capital should depend on the \emph{position} that is added to a portfolio and \emph{not} on the portfolio composition. This requirement implied a \emph{ratings} based approach to capital management. The second requirement is that the total cost of capital is given by the sum of the capital costs of the individual positions. The second requirement has been traditionally implemented by resorting to an assumption of infinite-granularity and to the use of the 1-factor Gaussian copula. This approach features a global single systemic risk factor. Within this framework the regulatory capital of a certain position depends ultimately on the rating of the position and on the capital cost associated with the rating class. Since the latter has been time invariant, the Basel regulatory capital of a portfolio has been a quite \emph{static} measure by construction. For a critic see for example \cite{Blundell}.

\subsubsection*{Economic Capital}
The economic capital of a portfolio, on the other hand, is supposed to depend on assumptions concerning the economic activity and conditions at the time of calculation. Additionally, the capital associated to a certain position depends on the specific portfolio in which the position is located. This point becomes much more relevant when the instrument in the portfolio is a securitization note. \emph{Asset backed securities} (ABS) have been in the market for at least 25 years, and the origin of that asset class has its roots in the US, where it has appeared as an attempt to address part of the economic problems caused by mortgages (see e.g. \cite{Kothari06}). The economic capital of a portfolio depends on the shape of the tail of the loss distribution, and this shape is very much affected by the portfolio assets \emph{correlations} (or, better, dependence; here we will use the word "correlation" also to denote the more general and well-posed notions of statistical dependence). The concept of correlation and its calibration depends significantly on the adopted \emph{assumptions} for the determination of the loss distribution~\footnote{The details are out of the scope of this paper. We refer to 	~\cite{GGTheArt2009} and \cite{BrigoPallaTorreWiley} for details.}.

\subsubsection*{Contingent Capital Bonds (CoCo's)}
As a consequence of the credit crisis, very large financial institutions are still under-capitalized at a time when economic development is most needed. To solve the issues of low capital levels~\footnote{The use of taxpayer money to bail out the financial system has become, expectedly, a matter of very intense debate.} the industry has recently proposed a new concept of capital called \emph{Contingent Capital} (\emph{Coco}'s). This instrument has a hybrid nature and is similar to a convertible bond, with some relevant caveats. In general a Coco is a form of capital that has a hybrid format: the Coco is a bond that may be converted into equity when or if a certain event happens. Typically, the event is related to a capital ratio falling below a predetermined threshold. The event, always related to the capital situation of the issuing institution, triggers the conversion of the bond into equity. In order to compensate the bond-holder for the risk of a conversion at a possible distressed time, the Coco bond would pay a more generous coupon than a similar bond without Coco's features. A crucial question is "how much more generous"?

\subsubsection*{CoCo pricing, default models and conversion}
A very important issue is then what would be the price that would cover that risk and how it would be determined. Coco's are driven by conversion, but conversion in turn may be driven by credit quality and default risk, so that we need to consider a model for the credit and the default time of the issuing institution. Generally speaking, there are two distinct ways to model the default process. 

The first default modeling area is the one we will adopt in the present work. This is known as the \emph{firm value} models area and it is generally attributed to~\cite{Merton1974} and \cite{BC76}, and underpins the Black-Scholes formula. In these models a company defaults when a latent variable, the firm asset value, breaches some barrier, typically the debt value and safety covenants. One then needs assumptions for the asset value process and for the capital structure of the company. Since the original work of Merton and Black and Cox there have been many extensions to it, and we refer to~\cite{BieleckiRutkowski2002} and the references therein for details. In this paper, we will use the \emph{analytically tractable first passage} model (AT1P)\footnote{The model has been developed by~\cite{BrigoT2004} for pricing Equity Swap under counterpart risk and by~\cite{BMT2009} for the same problem while analyzing Lehman Brothers.} developed by~\cite{BrigoT2004} to price a contingent capital bond. The model is calibrated to the implied risk neutral probabilities extracted from the CDS market by the use of analytical formulas for barrier options. 

The second default modeling area is termed "reduced form models" or "intensity models", see for example the original work of~\cite{Duffie1997}. In this case one renounces modeling the default from an economic point of view and just assumes that the default time is the first jump of a time-inhomogeneous Poisson process, where the intensities of default (or instantaneous credit spreads) are usually calibrated to prices of the CDS market. This has been the preferred approach by market practitioners interested in fitting (by construction) the implied risk neutral probability of default from the CDS market. In this context, the market is often assuming that the intensity is deterministic and that there is no credit spread volatility. This assumption is completely unrealistic in the light of the sizes of both implied and historical volatility for CDS, see for example \cite{Brigo2005}. Furthermore, when coupled with static copula functions for default correlation, the lack of spread volatility may cause counterintuitive and dangerous patterns and results in counterparty risk modeling, see for example \cite{BrigoChourdakis}.  
It is possible to include credit spread volatility in a tractable way while preserving CDS calibration. Indeed, models with credit spread volatility can still be tractable even for CDS options and can calibrate the CDS market quickly and precisely, see for example the works \cite{BrigoAlfonsi05} and \cite{BrigoElbachir10}.

The reason why we choose the first family of models is due to its appeal in terms of linking model features (a default barrier) with economic content (asset returns determine financial conditions). In intensity models default strikes out of the blue, following a totally unpredictable exponential random time, that is hard to connect to economic variables. In firm value models instead default is the result of a process involving firm value and debt, so that a clear link with economics is given. 

We will postulate a simple relationship between the conversion trigger and the Asset/Equity Ratio (AER), a quantity available in our firm value model. We then improve the relationship by allowing a given amount of de-correlation between the  conversion trigger and the AER.

\subsubsection*{This paper and earlier literature}


Given Capital concerns in banks, there has been much interest in CoCos in the literature. We may roughly classify previous research in two areas.
The first area deals with the pricing and risk management of CoCos instruments while the second area deals with investigating the impact of the introduction of CoCos on the optimal capital structure of the issuer. This paper lies in the first area. 

Concerning the second area of research, \cite{Barucci2011} extends an approach originally proposed by \cite{Goldstein2001}
where the underlying state variable is the claim to earnings before interest and taxes (EBIT) and where straight debt, contingent capital, equity and bankruptcy costs are modeled as perpetual assets, thus allowing for closed-form valuation formulas. 
The default barrier level, the trigger level and the optimal capital structure are then calculated such as to maximize the equity value and the net value of the company. Again, the paper  \cite{Barucci2011} shows as the optimal capital structure strategy and the bankruptcy costs can change depending on the introduction of CoCos and on the type of trigger event and rule for equity conversion that are adopted. 

Going back to the first area of pricing and risk management of CoCos, we refer the reader to \cite{Wilkens2012} and references therein for a fairly comprehensive review of the three main previous approaches to CoCo bonds valuation, which we may briefly describe as structural-default based, reduced-form based and equity modeling based.

The Structural-default approach models the issuer default via the first passage time of the firm value process through some threshold barrier associated with debt and safety covenants. The application of such models to CoCos pricing relies on the same idea and considers two different thresholds, the first one triggering equity conversion and the second one triggering default. In reduced-form based models, on the other hand, the trigger time is the first arrival time of a Poisson process with constant intensity, and the relevant intensity should be calibrated to the probability that the equity of the issuer falls below a certain threshold. A possible issue in reduced form models may be to model consistently equity conversion and default stopping times, since there is no economics guiding principle helping with coordination of the two times. 

Equity-based models decompose the CoCo payoff in terms of plain bonds, knock-in equity forward and a strip of knock-out equity puts. The knock-in and knock-out equity levels substitute the capital ratio trigger. For an introduction to CoCos and for a discussion of the different approaches to valuation see \cite{Schoutens2011}.

To the best of our knowledge, none of the existing approaches has considered the following three features explicitly and at the same time: 

\begin{itemize} 

\item Bond-Equity Conversion time, 

\item Equity price at Conversion, and 

\item Default time.

\end{itemize}

Without one or more of these three features it is not possible to calibrate a model to credit markets and equity market at the same time. This implies that one is discarding important information. The set of instruments to which the CoCo pricing model is calibrated affects the set of considered risk factors and, hence, hedging strategies. Models able to only capture the credit-component of this hybrid instrument, will only allow to hedge it through credit derivatives, while models able only to handle the equity component will only allow to use equity instruments. A CoCo bond, however, could react to information coming from both markets. 

This paper adds to the literature in important ways. Firstly, we show how to use the AT1P firm value model to price Coco's. Secondly, we describe a new calibration procedure, how to implement it, and finally we show the numerical results of our pricing approach, comparing our outputs with market quotes for a traded instrument. Thirdly, we analyze stress tests on our framework to see the impact on prices during periods of financial distress.

The paper is organized as follows. In section~\ref{sec:Cocos} we describe the Coco bond instrument and its use in practice, inclusive of the motivation underlying its creation. The AT1P model used in our pricing approach is summarized in section~\ref{sec:AT1P} while the discussion on its calibration is shown in section~\ref{sec:Calibration}. Results and a related discussion are presented in section~\ref{sec:Results}. Finally, in section~\ref{sec:Conclusions} we summarize our conclusions.

\section{Contingent Capital (Coco's)}
\label{sec:Cocos}
As mentioned in Section \ref{sec:Introduction}, Coco's is a new form of capital. They have been created with the main objective of increasing capital levels of systemically important large banks. These instruments' equity like feature may be attractive for banks in their effort to comply with the new regulatory capital rules required by the Basel III framework. 

Generally speaking, the Coco instrument takes the form of a convertible bond with the important caveat that the conversion trigger event is a bank capital ratio falling below a certain level known at the issuance of the instrument. In case the event is triggered, the coco bond will be converted into equity at a certain conversion price.  More precisely, in case the core capital ratio fell below a certain level, the financial institution would issue new shares (equity) to be bought at a certain contract pre-determined price. The way the conversion price is determined is specified in the contract and is an important part of the CoCo bond features.

\subsection{Conversion price determined at conversion date}

Indeed, an important feature of a Coco is its \emph{conversion price} (CP). There are two important and contrasting possibilities underlying the CP. In the first case the conversion price is determined at the conversion date. If so, the price might be far too low, possibly resulting in significant dilution for the shareholders. In that case, shareholders may certainly be tempted to sell their equity holdings as soon as they sense an increase in risk of the capital ratio reaching the barrier (triggering conversion). This ~\emph{rush to sell} may put the stock into a downward spiral that will potentially be self fulfilling. This approach, however, is quite convenient to the bond holders, given that they will be acquiring the shares at distressed prices. One can say that the approach will affect bond holders less at the expense of the equity holders. Depending on the equity price at the time of the conversion, the bond holders might become a quite large portion of the institution shareholders. Furthermore, depending on the created amount of equity, this might be seen as good news for the stock. 

\subsection{Conversion price fixed at the Coco inception}

If the conversion price, on the other hand, is pre-determined, there are different considerations to make. To begin with, it is important to understand what a reasonable initial value for the conversion price would be, in order to fix it at inception. 
In this sense, if the conversion price is determined as the spot price at the issuance date of the Coco, then the bond holder will be acquiring newly issued shares for a price that might be much higher than the actual equity price at the moment the conversion trigger kicks in. In this case, the dilution in shares will be relatively small while the losses for the bond holders will be large. The bond holder might request a higher coupon, given the risk of having to acquire equity for a considerably higher value than the market price (in case of conversion). This might seem a good solution for the equity holders, at the cost of penalizing the bond holders. The latter might be tempted to short the stock or buy put options. Bond holders will do so especially if they sense conversion as unavoidable and intend to hedge their losses at that event. Again there is a high possibility of downside pressure on the stock and of a self fulfilling "downward spiral". 


A different possibility would be to align the conversion price with the percentage in fall associated with the conversion trigger. That is, if the reduction trigger in capital amounts to say 40\%, then the equity conversion trigger would also amounts to a 40\% reduction in the spot equity price with respect to the known equity price at inception. From the equity holder perspective this might be a less radical solution than the first one, but in any case would still imply a considerable dilution effect.

\subsection{Transparency and inputs}
A second important point on Coco's has to do with \emph{transparency}. The instrument is strongly dependent on the capital ratios of a financial institution. Essentially, for pricing purposes, one needs to have access to the portfolio composition of the issuing institution. That is, one needs to have access to information on the portfolio compositions of the financial institution, or at least its aggregate percentage distribution in terms of sectors (and regions) and types of instruments. A portfolio composition however is a very hard information to obtain. In the next section we will present a model that does not depend on the portfolio composition while depending on the path of the equity return of the company.

\section{Default Models: From Merton to AT1P}
\label{sec:AT1P}


There are currently two main approaches to model default. The first approach is known as firm value, structural or equity based models and is inspired on the classic work of~\cite{Merton1974} and~\cite{BC76}, both building on the original framework of \cite{BS73}. The second approach is known as intensity or reduced form models, and is inspired by the seminal work of~\cite{Duffie1997}. In this work we will concentrate on Merton like models and we refer to~\cite{GGTheArt2009} and the references therein for details, see also \cite{BieleckiRutkowski2002} and \cite{brigocvabook}.

In the following we will use different notations for functions of time and stochastic processes, the former being denoted by $X(t)$ and the latter by $X_t$. Vectors will be indicated by boldface letters.

In structural models default occurs when a latent variable, the asset
value, breaches a barrier, typically the debt book value of the company. One
needs an assumption for the asset value process and an assumption for the capital
structure of a company. Let us consider a probability space $(\Omega,\mathcal F, \mathbb P)$ with a standard Brownian Motion defined on it, $W^P_t$, and the filtration generated by the Brownian motion, $\{\mathcal F_t\}_{t\geq 0}$.  Denote also by $V_t$ the value of the company, $S_t$ its equity price
and $B_t$ the value of its outstanding debt at time t. Additionally $D$ is the par or
notional value of the debt at maturity $T$. 

\subsection{The Merton and Black \& Cox models}
The Merton model only checks for default at the final debt maturity $T$. The default time is the maturity if the firm value is below the debt par level $D$, and is time infinity (meaning no default) in the other case.

Formally, the default time is defined as

\[ \tau = T  1_{\{ V_T < D \}} + \infty  1_{\{ V_T \ge D \}}, \]
while the value $V_t$ of the company is given by

\beq
{V}_t = {S}_t + {B}_t,
\label{eq:Vt}
\eeq
whereas, at maturity $T$, outstanding debt, firm value  and debt at maturity are related by $B_T = \min (D, V_T) = D - (D-V_T)^+$ so that
\begin{equation}
S_T = (V_T-D)^+.
\label{eq:St}
\end{equation}
Under Merton's assumptions the market is composed by $V_t$ and by a money-market account, $\beta(t) = \beta(0) e^{rt}$. $V_t$  is a Geometric Brownian Motion so that its dynamics is described by the following stochastic differential equation

\begin{equation*}
\dif{V_t} = \mu{V_t} \dif t + \sigma {V_t} \dif W_t^P,
\end{equation*}
where $\mu$ is the drift and $\sigma$ is the volatility. As usual $V_t$ is expressed as Geometric Brownian Motion in the equivalent risk neutral martingale measure $\mathbb Q$, so that 
\begin{equation}
\dif{V_t} = r {V_t} \dif t + \sigma {V_t} \dif W_t^Q,
\label{eq:dV}
\end{equation}
where $W^Q_t$ is a Brownian Motion in the probability space $(\Omega,\mathcal F, \mathbb Q)$.

In the original Merton model the value of the company should not fall below
the outstanding debt at maturity. In this case we have from  \eqref{eq:St} and   \eqref{eq:dV} that the value of the equity of a company at maturity T is given by the Black Scholes formula.

An important limitation of the Merton model is the assumption that default may only occur at final maturity $T$. This limitation was addressed first in \cite{BC76}, where early default is introduced, resulting from the breaching of safety covenants. Safety covenants are modeled according to a continuous time default barrier $H(t)$.   Formally, the hitting-time of the firm value process on the barrier $H$ is $\tau$. 
\[ \tau = \inf\{ t \ge 0:\ V_t \le H(t) \}, \ \ \inf \Phi := +\infty\]

Notice that $\tau$ is a random time with respect to $\{\mathcal F_t\}_{t\geq 0}$.
The Black Cox model improves the Merton model but still maintains limitations. The shape of the default barrier can only be exponential, and the parameters $r,q$ and $\sigma$ need to be constant. The next model we consider removes such limitation and provide the structural models with enough flexibility to compete with the reduced form models in the single name calibration space.

\subsection{AT1P model: calibration to the whole CDS term structure}
\label{subsec:calibration}

\cite{BrigoT2004} have extended the original Merton and Black Cox setting in two important ways:
\begin{enumerate}[i)]
\item by considering a deterministic non-flat barrier;
\item by allowing for a time-dependent volatility for the firm value process.
\end{enumerate}	

This approach has recently been used by~\cite{BMT2009} for pricing Lehman equity swaps taking into account counterparty risk. Notice also that these authors have considered random default barriers associated with misreporting and risk of fraud. The related research is summarized for example in~\cite{BMT2009}, where the random barrier framework is applied to Lehman's default. Here we stick to the first version of the model, having deterministic non-flat barriers.  

The reason for imposing a particular shape for the barrier is to bring feasibility / practicality to the calibration and pricing process. That is, the survival probabilities of the firm can be recovered using closed form formulas. Additionally, it is possible to consider different maturities for the outstanding debt of the firm. Finally, the main achievement of the AT1P generalization relies in the possibility of calibrating the model to the whole CDS term structure of the firm, as illustrated for Parmalat and Lehman in the above references.

We present here a formulation of the model in the simplified setting in which dividends and the risk free short rate are constant:
\begin{subequations}
\begin{align}
\mbox{Firm-value process}: & & \dif V_t &=(r-q)V_t  \dif t + \sigma(t)V_t \dif W^Q_t,\label{eqn:AT1P_proc}\\
\mbox{Barrier} :& & \hat H(t) &= H e^{(r-q)t- b \int_0^t \sigma^2(s) \dif s},\label{eqn:AT1P_barrier}\\
\mbox{Survival probability}:& & \mathbb Q(\tau > T) & =  \Phi\left(d_1 \right) - \left( \frac{H}{V_0} \right)^{2b-1}  \Phi\left(d_2\right),\label{eqn:AT1P_surv} \\
 & & d_1 &:=  \frac{\log \frac{V_0}{H}+ \frac{2b-1}{2} \int_0^T \sigma(s)^2 \dif s} {\left(\int_0^T \sigma(s)^2 \ \dif s\right)^{1/2}}, \nonumber\\
&  &d_2 & := d_1 -  \frac{2 \log \frac{V_0}{H}} {\left(\int_0^T \sigma(s)^2 \dif s\right)^{1/2}}. \nonumber\\
& & \tau &= \inf\left\{ t \geq 0 \ \mbox{s.t.}\  V_t \leq \hat H(t)\right\}
\end{align}\label{eqn:AT1P}
\end{subequations}


The model is easily generalized to time dependent $r$ and $q$, so as to be able to fit the term structure of interest rates and safety covenants, see~\cite{BrigoT2004} for the fully general model.  
In the following we will make the same choices proposed in~\cite{BrigoT2004} about $\sigma(\cdot)$, using a piecewise constant function to model the evolution of the firm-value's volatility. We will partition the time-domain of the volatility function in a number of time-steps equal to the number ($n$) of CDS contracts used for calibration. Consider the set of model parameters $\vet p := (B,H,\sigma_1,...,\sigma_n)$. 


In the next section we are showing how~\eqref{eqn:AT1P_surv} is used in the pricing of CDS's. 




\subsubsection{CDS pricing}

In this section the pricing formula for a standard CDS is derived. Let us assume for simplicity deterministic interest rates, and that the survival probabilities are given by a closed form formula as in~\eqref{eqn:AT1P_surv}. 


Consider a CDS contract in which protection is sold on a reference entity whose recovery rate is $R$. The contract starts in $T_0$, has maturity $T_N$, an annualized premium rate $S$ and premium leg payments to be done on a time-grid $T_1,T_2,\dots,T_N$, typically quarterly. The contract discounted payoff at time $t<T_0$ is given by
\begin{equation}
\Pi^{T_0,T_N}(t; S) = \sum_{i: T_i > t} \left( S \left[\frac{\beta(t)}{\beta(T_i)}  \alpha_i \mathbbm 1_{\{\tau>T_i\}}  + \frac{\beta(t)}{\beta(\tau)} \alpha_d \mathbbm 1_{\{T_{i-1}<\tau\leq T_i\}}\right]-  \frac{\beta(t)}{\beta(\tau)} (1-R)\mathbbm 1_{\{T_{i-1}<\tau\leq T_i\}}\right), 
\label{eqn:CDS}
\end{equation}
where $\alpha_i := (T_i-T_{i-1})$, $\alpha_d := (\tau-T_d)$ and $T_d$ is the last payment date before default,  $\mathbbm 1_{\{\tau>T_i\}}$ and $\mathbbm 1_{\{T_{i-1}<\tau\leq T_i\}}$ are the indicators for default after $T_i$ and between $T_{i-1}$ and $T_i$ respectively.

The price of the CDS at time $t$ is  the expected value of the above discounted payoff calculated using the measure $\mathbb Q$ and conditional on $\mathcal F_t$. We denote the price by $\mathbb E_t\left[\Pi^{T_0,T_N}(t; S)\right]$. Notice that only the expected value of the first term can be straightforwardly evaluated (using the probabilities given in~\ref{eqn:AT1P_surv}). The second term that accounts for the accrual between a payment date and the default time, and the default leg, can be evaluated precisely as Stieltjes integrals (see for example \cite{BrigoT2004}) and then trivially discretized for practical implementation. For example, the accrual can be approximated by half the full premium due to be paid at the end of each period. With respect to the protection leg, we use the premium leg time-grid as a discretization grid for the integral.  These two assumptions work well when credit spread are not too large, but need to be handled carefully in case of large spreads, where a more refined grid may be utilized. One obtains
\begin{multline}
\mathbb E_t\left[ \Pi^{T_0,T_N}(t; S)\right] =S\sum_{i: T_i > t} \frac{\beta(t)}{\beta(T_i)}\alpha_i\left( \mathbb P(\tau>T_i)+\frac 1 2 \left( {\mathbb Q}_t (\tau>T_{i-1}) - {\mathbb Q}_t( \tau>T_i) \right) \right)  \\
 - (1-R) \sum_i \frac{\beta(t)}{\beta(T_i)}\left( {\mathbb Q}_t(\tau>T_{i-1}) - {\mathbb Q}_t (\tau>T_i)  \right).
\end{multline}

The CDS par spread $S^T$ is the spread for the today-issued maturity $T$ contract such that the value of the contract is null. Once it is expressed as a function of the model parameters ($\vet p$) one uses a series of market quoted spreads in the calibration of the model (that is, finding the model parameters that match market observed CDS spreads). Assume $S^{T_0,T_N}(0)$ as the CDS spread at time $0$ for an insurance contract spanning quarterly dates from $T_0\ge 0$ up to $T_N$. The model spreads are such that   $\mathbb E_0\left[ \Pi^{T_0,T_N}\left(0; S^{T_0,T_N}(0)\right)\right] = 0$, that is

\begin{equation}	
S^{T_0,T_N}(0) = \frac{(1-R) \sum_i \frac{\beta(0)}{\beta(T_i)}\left( \mathbb Q(\tau>T_{i-1}) - \mathbb Q (\tau>T_i)  \right)}{\sum_i \frac{\beta(0)}{\beta(T_i)}\alpha_i\left( \mathbb Q(\tau>T_i)+\frac 1 2 \left( \mathbb Q (\tau>T_{i-1}) - \mathbb Q( \tau>T_i) \right) \right) }
\end{equation}•
and $\mathbb Q(\tau>T_a)$ can be evaluated using~\eqref{eqn:AT1P_surv}.

\begin{myDef}[Calibration procedure]\label{def:calibration}
In general we will mean the following problem by referring to calibration procedure. Let us consider a state space $\mathcal X$, market observable values $\boldsymbol \phi^M = (\phi^M_1,\dots,\phi^M_{N_M})$ and their correspondent model calculated quantities  $\boldsymbol \phi(\vet x) = (\phi_1(\vet x),\dots,\phi_{N_M}(\vet x))$ where $\vet x\in\mathcal X$ are  model parameters. The \emph{calibrated model parameters vector}  $\vet p$ is such that
\begin{equation*}
C(\vet p) = \min_{\vet x\in\mathcal X}C(\vet x)
\end{equation*}•
where $C(\vet x)$ is a distance between $\boldsymbol \phi^M$ and $\boldsymbol \phi (\vet x)$  (in the following $C(\cdot)$ can also be called \emph{cost function}). In those cases where a Newton-type algorithm will be used, we will refer to the $i$-th point in the iterative process as $\vet x^{(i)}$.

Through this work, we have used a two-steps approach:
\begin{enumerate}[i)]
\item we specified the minimization problem as 
\begin{equation*}
C(\vet x) = \sum_{i=1}^{N_M}w_i \left( \frac{\phi_i(\vet x) - \phi_i^M}{\phi_i^M} \right)^2,
\end{equation*}
with $w_i = \frac 1{N_M}$ and Levenberg-Marquardt algorithm to minimize $C(\cdot)$;
\item
we set the starting point, $\vet x^{(0)}$, for Levenberg-Marquardt algorithm by solving the same problem with a global search algorithm, in this case simulated annealing.
\end{enumerate}•
\end{myDef}

\subsubsection{On the parameters $B$ and $H$}
\label{sec:AT1P_results}
Let us consider CDS par spreads on junior debt issued by Lloyd's as of 15-Dec-2010. They are shown in Table \ref{tbl:quote}. 
\begin{table} 
\centering\begin{tabular}{| r |r @{.}l |}
\hline
  Term &
 \multicolumn{2}{c|}{CDS Rate (bps)}  \\[1ex]
\hline

1Y &
  347&9934 \\
2Y &
  373&1248 \\
3Y &
  396&6364 \\
4Y &
  417&8327 \\
5Y &
436&3855 \\
7Y &
  441&1132 \\
10Y &
  445&8688 \\
\hline
\end{tabular}\caption{CDS quotations on junior debt issued by Lloyds on 15-Dec-2010 (Source: Fitch Solutions.}\label{tbl:quote}
\end{table}

The number of parameters to describe $\sigma(\cdot)$ is equal to the number of CDS contracts included in the calibration. The volatility-linked parameters $\sigma_1,\dots,\sigma_n$ can provide enough flexibility to fit the ($n$) CDS par-spreads on their own, no matter the values of $B$ and $H$. This leaves us with two additional degrees of freedom. Table \ref{tbl:b0results} shows the results of four different calibrations. In all these four calibrations $B$ has been constrained to be equal to zero. The calibrations differ from each other due to different choices of the initial value for $H$.

\begin{table}
\centering\begin{tabular}{| l | r@{.}lr@{.}l |  r@{.}lr@{.}l | r@{.}lr@{.}l | r@{.}lr@{.}l |  }
\hline
  Term &
 \multicolumn{4}{|c|}{$H^{(0)} = 0.2$} &
    \multicolumn{4}{c|}{$H^{(0)} = 0.4$} &
 \multicolumn{4}{c|}{$H^{(0)} = 0.6$} &
    \multicolumn{4}{c|}{$H^{(0)} = 0.8$} \\
\cline{2-17}
 &  \multicolumn{2}{|c}{$\sigma$} &
    \multicolumn{2}{c|}{$\Delta S/S^M$} &
 \multicolumn{2}{c}{$\sigma$} &
 \multicolumn{2}{c|}{$\Delta S/S^M$} &
 \multicolumn{2}{c}{$\sigma$} &
 \multicolumn{2}{c|}{$\Delta S/S^M$} &
 \multicolumn{2}{c}{$\sigma$} &
   \multicolumn{2}{c|}{$\Delta S/S^M$} \\
 &  \multicolumn{2}{c}{} &
    \multicolumn{2}{c|}{$(10^{-14}\times)$} &
 \multicolumn{2}{c}{} &
 \multicolumn{2}{c|}{$(10^{-14}\times)$} &
 \multicolumn{2}{c}{} &
 \multicolumn{2}{c|}{$(10^{-14}\times)$} &
 \multicolumn{2}{c}{} &
   \multicolumn{2}{c|}{$(10^{-14}\times)$} \\
\hline
1Y&
  0&3019 &  0&0  & 
  0&2316 & -0&3775 & 
0&1737 & 0&3775 &
0&1330& -0&4219  \\
2Y &
   0&1171 &  0&1554  & 
  0&0938 & -0&0222 & 
0&0729 & -0&0888 &
0&0572 & -0&0666  \\
3Y &
  0&1992 &  0&1110  & 
  0&1558 & -0&0999 & 
0&1189 & -0&0777 &
0&0922 & -0&1998  \\
4Y &
   0&1886 &  -0&0222  & 
  0&1495 & -0&1110 & 
0&1154 & 0&0666 &
0&0902 & -0&0888  \\
5Y &
  0&2088 &  -0&0666  & 
  0&1663 &- 0&0333 & 
0&1289 & 0&0444 &
0&1011 & -0&0333  \\
7Y &
  0&1946 &  -0&0222  & 
  0&1565 & -0&0666 & 
0&1225 & 0&0444 &
0&0967& 0&0222  \\
10Y &
   0&2182 &  0&0222  & 
  0&1780 & -0&1221 & 
0&1653 & 0&0222 &
0&1126& -0&0333  \\
\hline
\end{tabular} \caption{Calibrated volatilities and correspondent CDS spread relative errors under different initial guesses for $H$. Here $\Delta S$ is the difference between the model spreads and the market spreads $S^M$.} \label{tbl:b0results}
\end{table}

Table \ref{tbl:b0results} shows that the four calibration problems converged with similarly good fit to the market observed spreads. That is the four calibrations result in a model satisfactorily near to the benchmark market quantities. However, different starting points for the barrier level resulted  in different calibrated values both for sigmas and for $H$ (see Table \ref{tbl:H} for values for calibrated $H$), due to the over-parametrization for the model. This issue will be addressed later on, when we use the AT1P model to price a CoCo bond.
%

\begin{table} 
\centering\begin{tabular}{| l | r@{.}l | r@{.}l  | r@{.}l | r@{.}l |}
\hline
   &
 \multicolumn{2}{|c|}{$H^{(0)} = 0.2$}  &
 \multicolumn{2}{c|}{$H^{(0)} = 0.4$} &
 \multicolumn{2}{c|}{$H^{(0)} = 0.6$}&
 \multicolumn{2}{c|}{$H^{(0)} = 0.8$}\\[1ex]
\hline

Calibrated $H$ &
  0&5584 & 0&6431 & 0&7205 & 0&7794 \\
\hline
\end{tabular}\caption{Calibrated values for $H$ correspondent to four different starting points.}\label{tbl:H}
\end{table}

%
%
%
%

\section{Adapting AT1P to CoCos}
\label{sec:Calibration}

When dealing with a contingent conversion bond, in addition to the default time $\tau$ we need to model a second random time. This second time accounts for the conversion of the CoCo bond. Assume $\tau_c$ to be the (random) conversion time that is defined in the contract as
\begin{equation*}
\tau_c = \inf\{t\geq 0\  : \ c_t \le \bar c \},  \ \ \ \inf \Phi := \infty 
\end{equation*}
where $c_t$ is the \emph{regulatory capital} at time $t$ and $\bar c$ is the threshold level below which the CoCo bond is converted into equity. In the current set up, $c_t$ is an exogenous variable whose proxy is shown in section~\ref{sec:proxy}. An alternative approach would involve the direct modeling of $c_t$. This is left for further research.

\subsection{CoCo's Payoff}

Given that $\tau$ and $\tau_c$ are not independent, a natural question is whether a CoCo bond can default before being triggered. In this regard we can make some considerations:
\begin{itemize}
\item these instruments are designed specifically to avoid that, in case of troubles caused by capital issues, banks will need to go through expensive new stocks issuances;
\item at default $t=\tau$, it makes sense to have $c_\tau = 0$;
\item regulatory capital information is neither continuously (say daily in this context) updated, nor it is as transparent as liquidly traded quantities. This capital information is expected to be updated at most twice a year, on the balance account dates of June and December.
\end{itemize}

In practice the updating frequency of $c_t$ is quite low (usually twice a year) and it can certainly happen that default occurs between two updating dates. Let us assume for the time being that issuers and regulators alike will use the conversion option as a last resort \emph{before} any technical default\footnote{As it will become clear in section~\ref{sec:proxy} one does not need this assumption as it is in fact a \emph{consequence} of the proxy assumed for $c_t$ (see~\ref{eqn:c_proxy} for detail).}. The relation between $\tau$ and $\tau_c$, default and CoCo conversion time respectively, is then constrained by  

\begin{equation}
\mathbb P(\tau< \tau_c\ |\ \tau < T) = \mathbb Q(\tau< \tau_c\ |\ \tau < T) = 0. \label{eqn:Ptau}
\end{equation}

This implies that, before $T$, the default of the bond will always follow the conversion time. This way, since at conversion the payoff is converted into equity, the payoff of the CoCo bond depends explicitly on $\tau_c$ and equity only, with no explicit dependence on $\tau$. Of course equity will depend on $\tau$ implicitly. 

Consider now two times $t$ and $T$, namely the evaluation and the maturity dates respectively. Assume $\Pi_c(t,T)$ and $\Pi(t,T)$ to be respectively the discounted payoff of the CoCo bond and that of an identical risk-free bond without the conversion feature. Additionally, suppose $D(t,T) = \beta(t) / \beta(T)$ to be the risk free discount factor at time $t$ for the period $[t,T]$, and let $E_t$ be the value of the stock price at time $t$. We can write
\begin{equation}
\Pi_c(t,T) = \surv{c}{T} \Pi(t,T) + \mathbbm 1_{\{\tau_c\leq T\}}\left(\Pi(t,\tau_c) + N E_{\tau_c}D(t,\tau_c)\right), 
\label{eqn:firstPayoff}
\end{equation}
where the number of shares $N$ obtained in case of conversion is set at the CoCo issuance date in a way that $NE^\star =   \expV{\Pi_c(0,T)}{0}$. Here $E^\star$ denotes the stock conversion price, and~ \eqref{eqn:firstPayoff} can be rewritten as
\begin{equation}
\Pi_c(t,T) = \surv{c}{T} \Pi(t,T) + \mathbbm 1_{\{\tau_c\leq T\}} \left( \Pi(t,\tau_c) + \expV{\Pi_c(0,T)}{0} \dfrac{ E_{\tau_c}}{E^\star}D(t,\tau_c)\right). \label{eqn:cocoPayoff}
\end{equation}

In passing we observe that the following analogy is possible with traditional CDS models. Usually the CoCo bond is issued at par, meaning that $ \expV{\Pi_c(0,T)}{0}=1$, so that 

\begin{equation}
\Pi_c(t,T) = \surv{c}{T} \Pi(t,T) + \mathbbm 1_{\{\tau_c\leq T\}} \left( \Pi(t,\tau_c) + \dfrac{ E_{\tau_c}}{E^\star}D(t,\tau_c)\right) \label{eqn:cocoPayoff2}.
\end{equation}
If one interprets the conversion time as a default time, $E_{\tau_c}/ E^\star$ can be thought as a recovery rate. Formula~\eqref{eqn:cocoPayoff2} suggests an alternative approach to the valuation of these instruments via a reduced-form approach. That is, we could use the first jump of a Poisson process as a model for the conversion time. A full analogy with the reduced-form approach used to price CDS contracts, would mean that 
\begin{enumerate}[i)]
\item
$\tau_c$ would no longer be a $\{\mathcal F_t\}_{t\geq0}$ stopping-time;
\item
 the intensity of the underlying Poisson process could be made random and linked with other variables (rates and equity/recovery), but the exponential component of the random time would be independent of all Brownian motions. This may lead to a weak dependency between conversion time and other market drivers.
\end{enumerate}
It is worth stressing that in the AT1P framework we can have strong dependency between $\tau_c$ and $E_{t}$ as they are driven by the same factor.

\subsection{A proxy for $c_t$}
\label{sec:proxy}

We estimated a proxy for $c_t$ in terms of total assets and total liabilities as   
\begin{align*}
\mbox{Total Assets}\  A_t &\approx V_t;\\
\mbox{Total Liabilities}\ &\approx \hat H(t).
\end{align*}

Consider \emph{total equity} to be given by the difference between total asset and total liabilities. A common indicator of the leverage of a company is the ratio between total assets and total equity (also known as Asset/Equity Ratio (AER)). We assume  here that this ratio is the driver for the change in capital ratios. To have an idea of how the capital ratio behaves with respect to the AER, we selected from \emph{Fitch's Financial Delivery Service (FDS)}~\footnote{FDS is a proprietary financial database that includes banks and insurance companies.} all the end-of-year balance account dates. We filtered at these dates all the entities by their \emph{individual ratings}\footnote{The individual/viability rating is a concept developed by Fitch Ratings and we refer to the company's website for details.} and, for every individual rating class $i$ and for every end-of-year balance account date, we went through a cross sectional ordinary least square (OLS) estimation of the following statistical model
\begin{align}
c_i &= \alpha + \beta X_i + \epsilon_i,\label{eqn:statModel} \\
X_i &:= \frac{A_i}{A_i-L_i}. \nonumber
\end{align}
where $\epsilon$ is the regression residual term. 
This calibration step is a determining aspect in the methodology. Important points to keep in mind are the following. 
\begin{itemize}
\item First, \emph{individual ratings} do not assume that financial institutions are inherently protected by an underlying government support. 
\item Second, individual ratings are based on a database spanning more than 20 years of observations. This means at least three business cycles, and shows a history of the enormous impact and relevance of the information technology in the industry. \item Finally, although each financial institution might invest according to its own idiosyncratic strategy, we assume that financial institutions may be aggregated in homogeneous groups for capital purposes. That is, we are assuming that the \emph{individual ratings} capture the long term aspect of the institutions that are supposed to be reflected in the price of the instrument.
\end{itemize}

\begin{figure}
\centering
\subfloat[Scatter plot and OLS estimation of Tier 1 Capital Ratio against Asset/Equity ratio on December 2009 for class rating C. ]{\includegraphics[width = 0.4\textwidth]{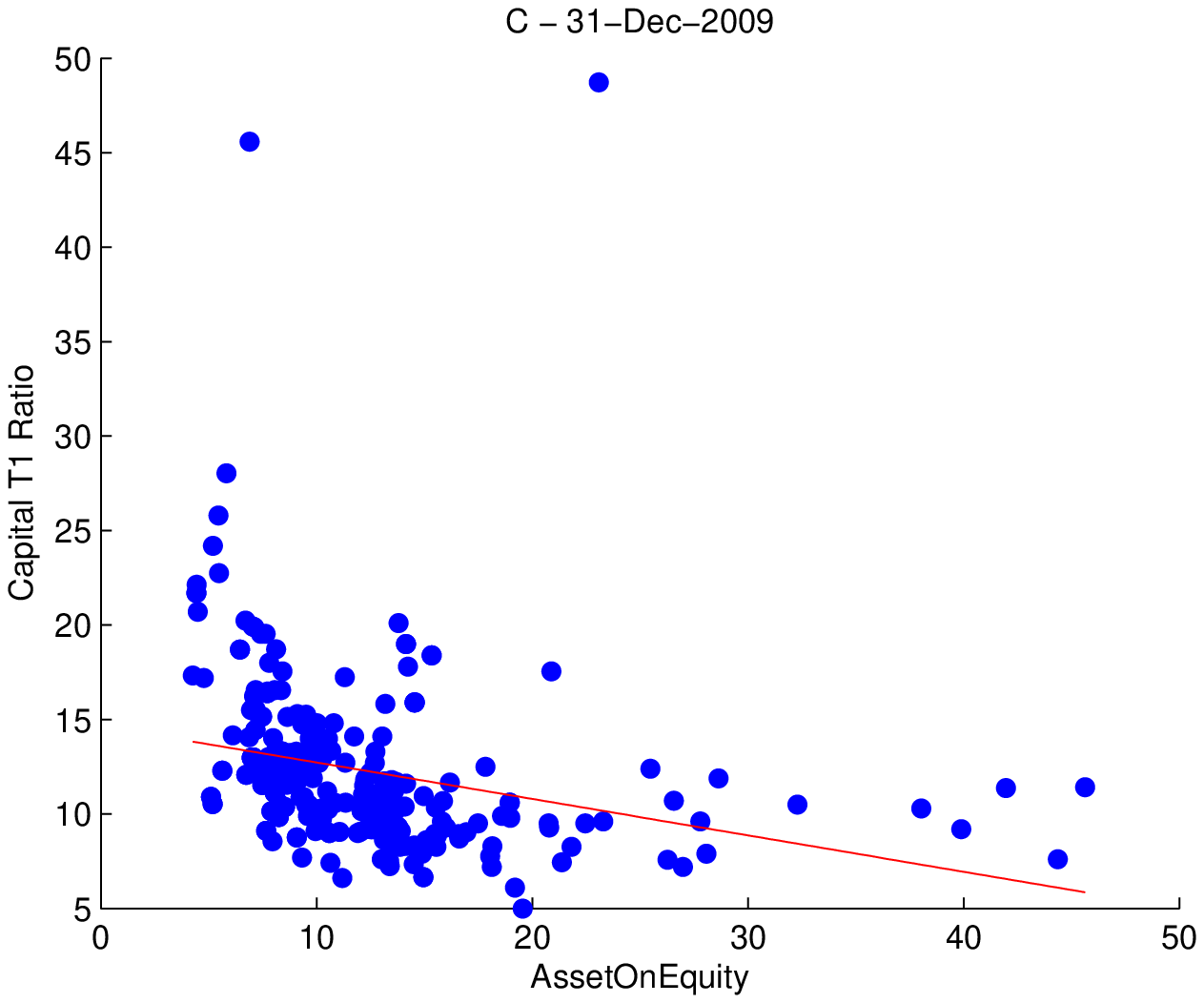}\label{fig:a}}
\subfloat[Evolution of $\alpha$ and $\beta$ for the C rating class]{\includegraphics[width = 0.45\textwidth]{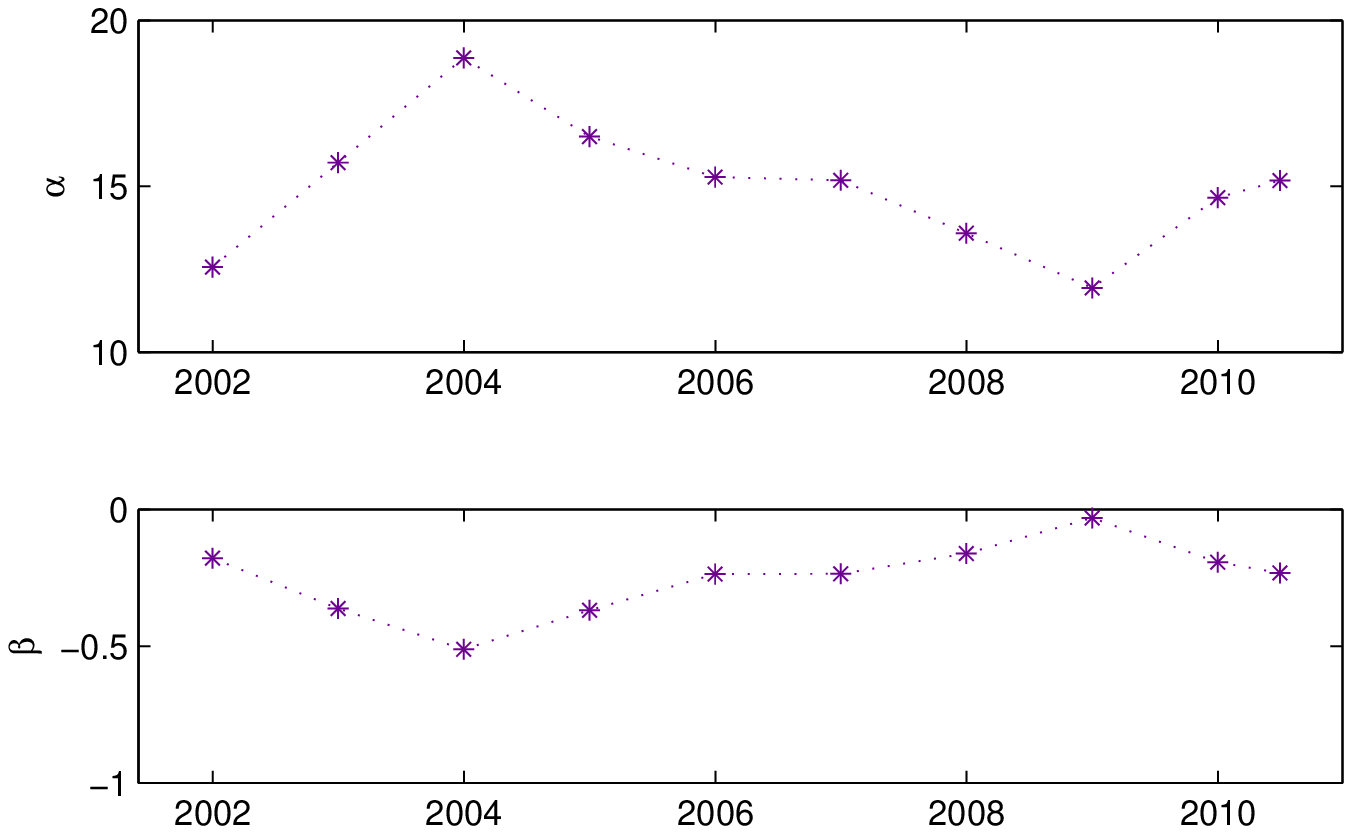}\label{fig:b}}
\caption{Regression results for the model in Eq. \eqref{eqn:statModel}}
\label{fig:statModel}
\end{figure}

In Figure \ref{fig:statModel} we show some results for the OLS estimation. In particular, Figure \ref{fig:a} contains the scatter plot and the simple linear regression for the balance account date of 31/12/09 and for the individual rating class C. Figure \ref{fig:b} shows the evolution, year by year, of parameters $\alpha$ and $\beta$ for the same rating class.

For every individual rating class, $\gamma$, we chose a simple average of the evolution over time of the two parameters $\alpha$ and $\beta$. Let us call the two averaged values $\bar \alpha_\gamma$ and $\bar \beta_\gamma$. It is possible to express our proxy for the regulatory capital ratio as $c_t = \hat c(V_t, \hat H(t); \gamma)$, with
\begin{equation}
\hat c(V_t, \hat H(t); \gamma) = \left\{ 
\begin{array}{lr}
\bar \alpha_\gamma + \bar \beta_\gamma\frac{V_t}{V_t-\hat H(t)} & V_t \geq \hat H(t), \\
0 & V_t < \hat H(t).
\end{array}•
\right. \label{eqn:c_proxy}
\end{equation}•

Notice that, once the model has been calibrated, Eq. \eqref{eqn:c_proxy} allows us to evaluate the price of a CoCo bond via a Monte Carlo simulation using the payoff as in Eq. \eqref{eqn:cocoPayoff2}.

The following proposition specifies under which conditions the proxy in~\eqref{eqn:c_proxy} implies the condition in~\eqref{eqn:Ptau}. 
\begin{prop}
Let us define $f(X_t) = \alpha + \beta X_t$ and assume  $f(X_0) >\bar c$ and $\beta<0$. Then the triggering event occurs before default, namely
\[ {\mathbb Q}(\tau_c < \tau) = 1 .\]
\begin{proof}
We have easily $X \rightarrow \infty$ as $V \rightarrow \hat H^+$ and, given that $\beta<0$, $f\rightarrow -\infty$ as $X\rightarrow\infty$. This means that in the limit of a default occurring we would have $f\rightarrow -\infty$. Observing that $f$ is a continuous function completes the proof, since $f$ would have to pass by $\bar c$ before reaching the default barrier.
\end{proof} 
\end{prop}

The relation between $c$ and $X$ for the individual rating C is shown in Figure \ref{fig:a}. In terms of our proxy we can infer from Figure \ref{fig:b} that the value of $\bar\beta_\gamma$ is negative. Eventually, as the proxy starts from a value higher than  the contractual trigger $\bar c$ it becomes relevant that we add it to the calibration procedure requiring that $\hat c(V_0, \hat H(0); \gamma)$ matches the last reported capital ratio . We will address this point in more detail in Section \ref{sec:calibration}.

\subsubsection{Decorrelating capital ratio and total asset}

We chose a setting which implies extreme values for the correlation between the capital ratio and the process AER. Indeed a straightforward calculation shows that
\begin{equation}
\corr (c_t^\gamma,X_t) = \sign (\beta).
\end{equation}• 
We can however obtain intermediate values for the correlation considering an alternative shape for the capital ratio. We propose 

\begin{equation} \label{eqn:newC}
\mathcal C_t^\gamma = \bar \alpha_\gamma + \bar \beta_\gamma \std(X_t) \left( \eta \frac{X_t}{\std (X_t)} + \sqrt{1-\eta^2} \epsilon_t^\gamma\right)
\end{equation}• 
where we've introduced 
\begin{itemize}
\item
a second random shock, independent of the Brownian motion underlying the total asset  $\epsilon_t^\gamma \sim \mathcal{N}(0,1) \quad \forall t$
\item
and a new parameter $\eta \in [0,1]$.
\end{itemize}•

Model \eqref{eqn:newC} satisfies the following two conditions
\begin{enumerate}[i)]
\item
Conservation of variance. The new capital ratio has the same variance as the old one
\begin{equation*}
\var (\mathcal C_t^\gamma) = \var (c_t^\gamma);
\end{equation*}•
\item
Generalization. The old capital ratio is a special case of the new one as
\begin{equation*}
\mathcal C_t^\gamma \rightarrow  c_t^\gamma \quad\mbox{ for } \eta \rightarrow 1.
\end{equation*}•
\end{enumerate}•

Furthermore, the new shape of the capital ratio provides an easy way to control its correlation with the AER process as it holds
\begin{equation}
\corr (\mathcal C_t^\gamma, X_t) = \eta \sign(\beta).
\end{equation}•
It is in view of the above equation that we will often refer to $\eta$ as  the correlation parameter. Finally, we will refer to the time step $\Delta t$ for checking the capital ratios as to the "sampling frequency".

\subsection{How to evaluate the stock-price in AT1P}
\label{sec:equity}

As explained in Section \ref{sec:Calibration}, in AT1P an entity can default at any time and not only at maturity. If we assume that the debt still has a clear single final maturity $T$ and that early default is given by safety covenants, it is not unreasonable to model equity as an option on the firm value with maturity $T$ that is killed if the default barrier is reached before $T$. See also the Equity chapter in  \cite{brigocvabook}. Namely, we calculate the stock price $E_t$ (in this framework) as a down-and-out European call option, that  is
\begin{equation}
E_t = \beta(t) \expV{\frac{\left(V_T-\hat H(T)\right)^+ \survInd{T}}{\beta(T)}}{t} = f(t,V_t).\label{eqn:fromV2S}
\end{equation}
We can use~\eqref{eqn:fromV2S}  to calculate the stock price both inside a simulation of~\eqref{eqn:cocoPayoff2} and in the calibration procedure. 

A closed form solution for the price of the option is given for example in \cite{Rapisarda2003}, see also the Equity chapter in \cite{brigocvabook}. 

The formula is as follows 

\begin{multline}
f =\frac{\beta(t)}{\beta(T)}\Biggl( V_t e^{\int_t^T(v(s)+\frac{\sigma(s)^2}{2})\dif s} \left(1 - \Phi(d_3)  \right) - \hat H(T)\left(  1-\Phi(d_4)\right) \\
- \hat H(t) \left(\frac{\hat H(t)}{V_t} \right)^{2B}  e^{\int_t^T(v(s)+\frac{\sigma(s)^2}{2})\dif s}\left( 1- \Phi(d_5) \right) + 
\hat H(T) \left(\frac{\hat H(t)}{V_t} \right)^{2B-1} \left( 1- \Phi(d_6) \right) 
\Biggr),\\
\end{multline}•
where  
\begin{align*}
v(t) &= r - q - \frac{\sigma(t)^2}{2},\\
d_3 &=  \frac{\left(\log \frac{\hat H(T)}{ H}\right)^+-\log\frac{V_t}{ H}-\int_t^T(v(s)+\sigma(s)^2)\dif s}{\left(\int_t^T \sigma(s)^2 \dif s\right)^{1/2}}, \nonumber \\
d_4 &=  \frac{\left(\log \frac{\hat H(T)}{ H}\right)^+-\log\frac{V_t}{ H}-\int_t^T(v(s))\dif s}{\left(\int_t^T \sigma(s)^2 \dif s\right)^{1/2}}, \nonumber \\
d_5 &=  \frac{\left(\log \frac{\hat H(T)}{ H}\right)^+-\log\frac{\hat H(t)^2}{H V_t}-\int_t^T(v(s)+\sigma(s)^2)\dif s}{\left(\int_t^T \sigma(s)^2 \dif s\right)^{1/2}}, \nonumber \\
d_6 &=  \frac{\left(\log \frac{\hat H(T)}{H}\right)^+-\log\frac{\hat H(t)^2}{H V_t}-\int_t^T(v(s))\dif s}{\left(\int_t^T \sigma(s)^2 \dif s\right)^{1/2}}. \nonumber \\
\end{align*}•
\subsubsection{Some remarks on the stock price evaluation}
In this section we will try to capture the effects of some of the AT1P model features in the stock price evaluation.

\begin{figure}
\centering
\subfloat[$V^{BS}_0\mapsto   \beta(t) \expV{\frac{\left(V^{BS}_T-K\right)^+}{\beta(T)}}{t}$  with \newline $K = \hat H(0)$. ]{\includegraphics[width = 0.45\textwidth]{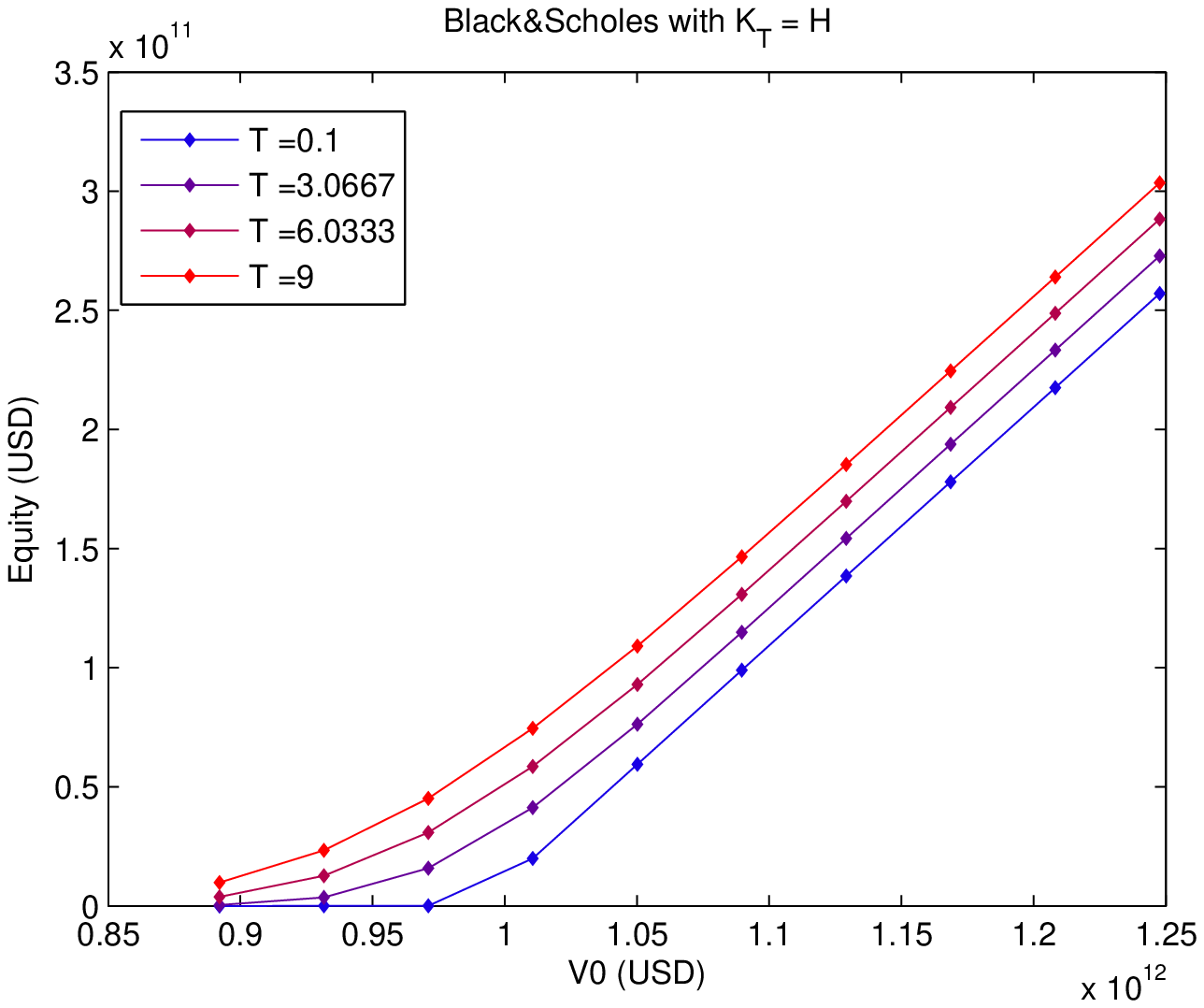}\label{fig:a}}
\subfloat[$V_0^{BS}\mapsto   \beta(t) \expV{\frac{\left(V^{BS}_T-K(T)\right)^+}{\beta(T)}}{t}$   with \newline $K(t) = \hat H(t)$.  ]{\includegraphics[width = 0.45\textwidth]{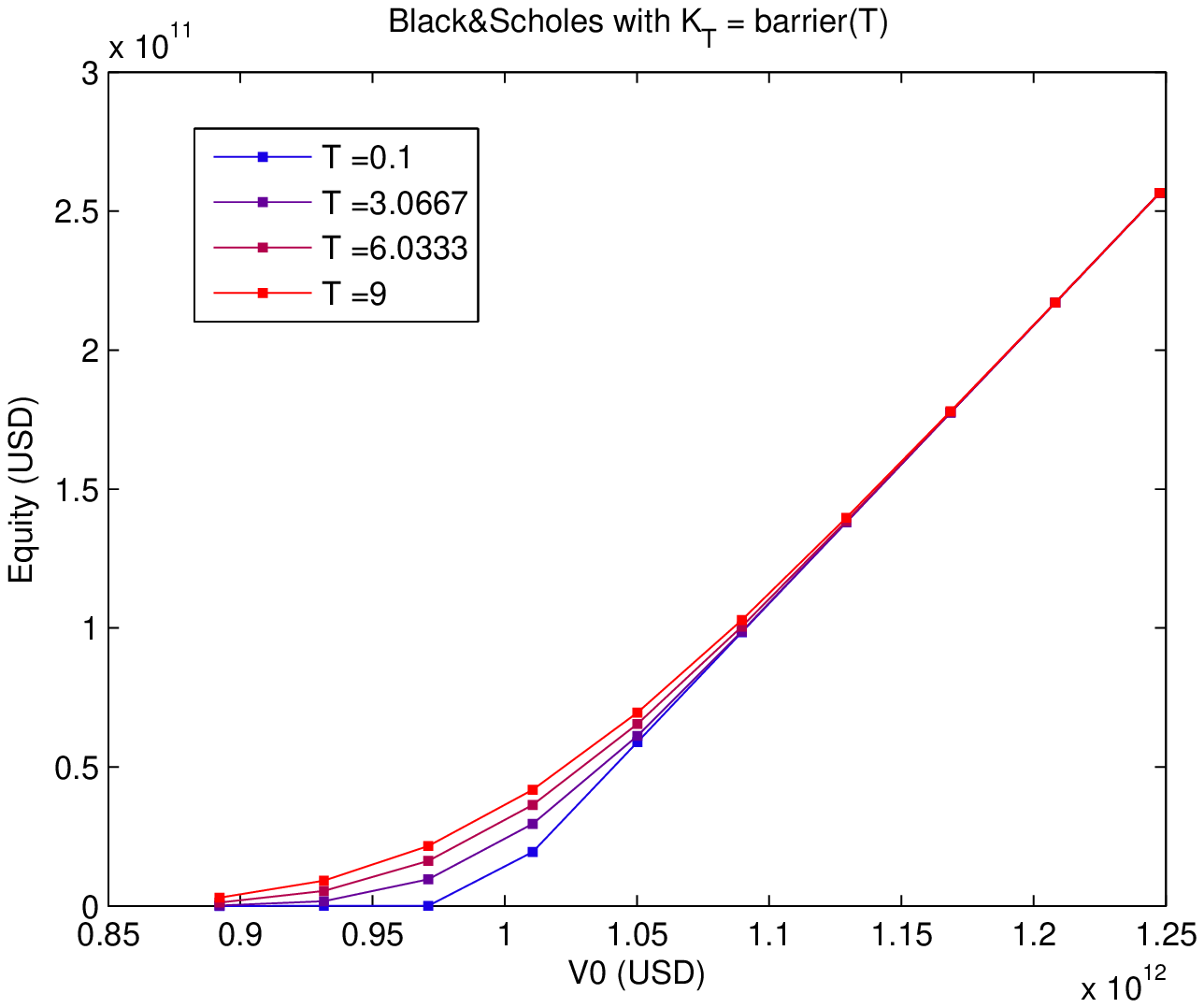}\label{fig:b}}\\
\subfloat[Equity spot price in AT1P. ]{\includegraphics[width = 0.45\textwidth]{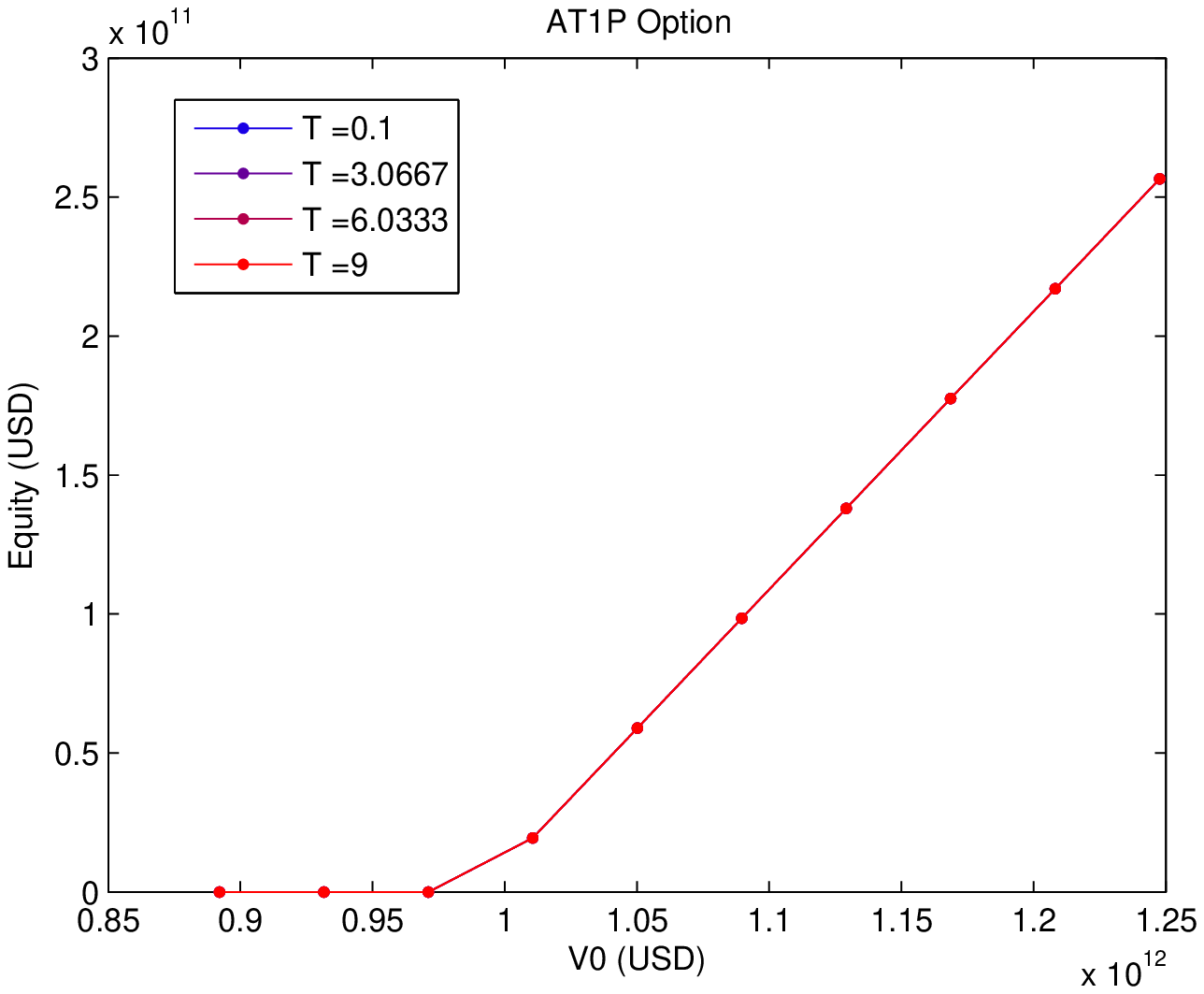}\label{fig:c}}
\subfloat[ $V_0\mapsto   \beta(t) \expV{\frac{\left(V_T-\hat H(T)\right)^+}{\beta(T)}}{t}$  ]{\includegraphics[width = 0.45\textwidth]{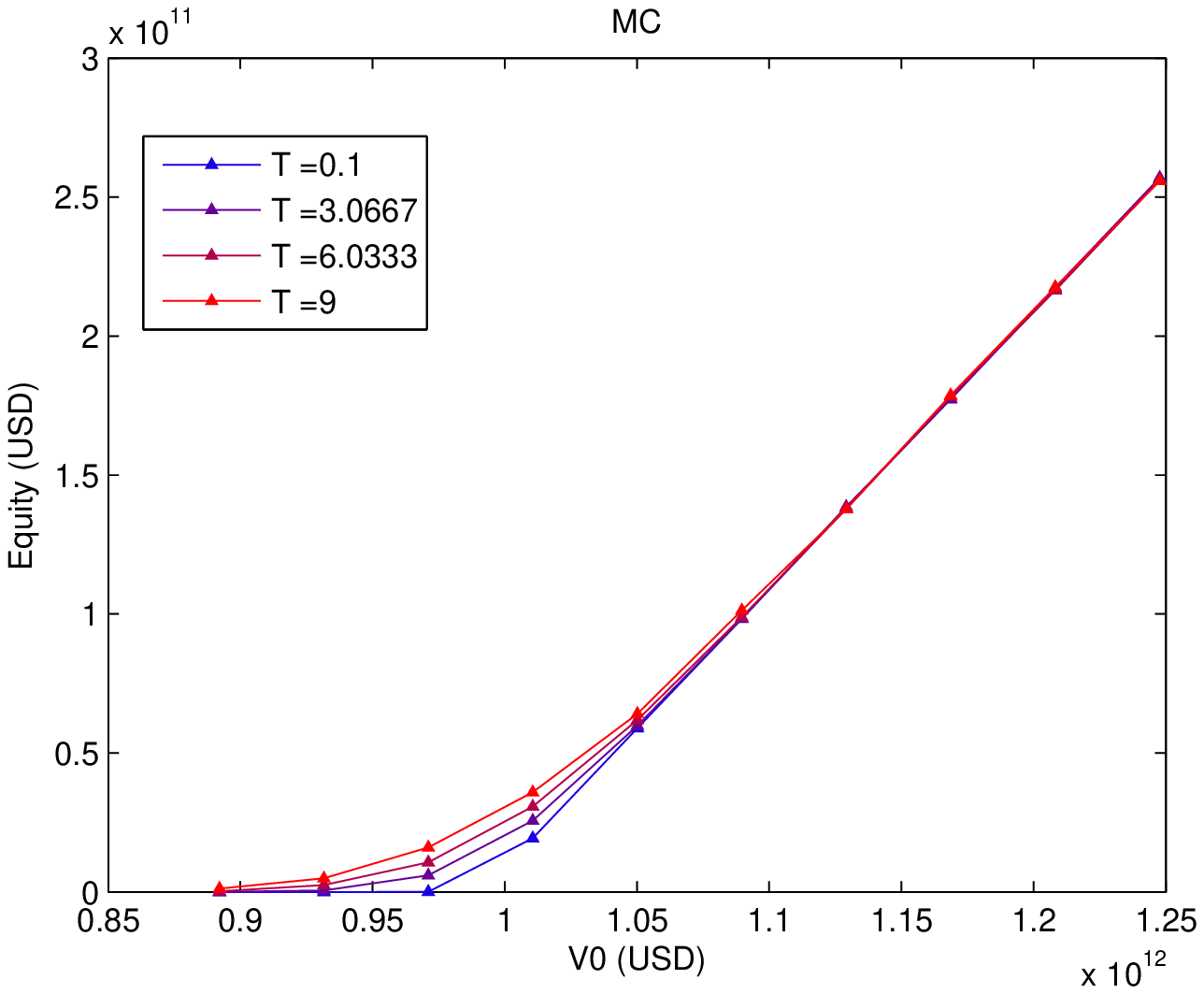}\label{fig:d}}
\caption{Comparison between using the AT1P model and the (widely spread) Black and Scholes formula. Top charts shows B\&S results, where  $\dif V^{BS}_t = r V^{BS}_t\dif t + \sigma_1 V^{BS}_t \dif W_t$ with $W$ a Brownian motion under $\mathbb{Q}$. Bottom charts show results where $V_t$ follows the process from \eqref{eqn:AT1P_proc} with $q =0$. }
\label{fig:equityPrice}
\end{figure}

We reported in  Figure~\ref{fig:equityPrice}  a series of plots illustrating a comparison between the  solution of~\eqref{eqn:fromV2S} 
and the evaluation of an European call option under the classic Black and Scholes framework.  We used  a set of parameters calibrated to the relevant market prices and such that $ B = 0$.

Observe from~\eqref{eqn:AT1P_barrier} that, in case $B=0$, the growth rate of $\hat H(\cdot)$ is the same as the one of the mean value of $V_t$.

Main points we wanted to highlight by showing Figure \ref{fig:equityPrice} are the following ones:
\begin{enumerate}[i)]
\item
Comparison between the plots in Figures \ref{fig:a} and \ref{fig:b}  shows the impact of considering a fixed strike $K$ against a time-dependent strike $K(t) = \hat H(t)$ that follows the safety covenants barrier. The latter results in a effect similar to considering a null drift for the underlying process;
\item
Comparison between plot in Figure \ref{fig:b} and the one in Figure \ref{fig:d} shows that the same behavior is attainable in the AT1P framework when we modify the payoff in  \eqref{eqn:fromV2S} by considering a plain call option rather than a down-and-out call option, that is by removing the indicator function inside the expectation\footnote{We resorted to a MonteCarlo engine to calculate the call prices in AT1P.};
\item
Comparison between charts in Figure \ref{fig:d} and Figure \ref{fig:c} shows that adding the down-and-out feature results in a sensible decrease of the diffusion effect, effectively making the option price fairly similar to a forward in the region where it is in-the-money.
\end{enumerate}•

It is particularly interesting to note that point iii) seems a confirmation of a result found in Chapter 8 in \cite{brigocvabook}.



\subsection{Calibration procedure for a CoCo Bond in AT1P}
\label{sec:calibration}
We followed the calibration procedure described in Section \ref{subsec:calibration} to determine the model's parameters. It is worth noticing that the model was calibrated not only to the quoted CDS spreads: indeed we used the results from Sections \ref{sec:proxy} and \ref{sec:equity} to include the flexibility of calibrating the model to the spot value of the stock price and to the most recent available regulatory capital ratio. The inclusion of these two additional market observables was needed to fix the indetermination in the values of the remaining model parameters (e.g. H),  shown by results in Tables~\ref{tbl:b0results} and \ref{tbl:H}.

As explained in Definition \ref{def:calibration}, we used in sequence simulated annealing and  Levenberg-Marquardt algorithms to perform the minimization needed for this calibration. 


In particular, considering $n$ quoted CDS par spreads for different maturities $(S^{T_1},\dots, S^{T_n})$, the last published tier 1  capital ratio , $c$, and the spot price of the equity $E_0$, the calibration works as follows:

\begin{enumerate}
\item
simulated annealing is used to set a starting point, $\vet x^{(0)}$, in the region  $\mathcal X :=  (0,5) \times (0,1)^{n+1}$. 
\item
starting from $\vet x^{(0)}$, we used  Levenberg-Marquardt to find the optimum point $\vet p$, in the state space $\mathcal X :=  \mathbb R_+ \times (0,1)  \times \mathbb R^n_+$, whereas we considered the set of market observables $ \phi^M = (S^{1Y},\dots,S^{10Y},c,E_0)$.
\end{enumerate}•

\section{Results}
\label{sec:Results}
In this section we will show numerical results for the 2-step calibration procedure described above. Additionally, we will illustrate the algorithm by using it to price a Lloyds issued CoCo bond. 	

Before we move forward an important point to keep in mind is the following. As we have seen in section~\ref{sec:AT1P}, the AT1P model is a \emph{Gaussian} based  model and the advantage of it is certainly mathematical tractability. This happens in the form of analytical formulas that make the model very handy for the calibration process. However, this model also suffers from a number of problems. The basic AT1P suffers from lack of short term credit spreads, common to more basic firm value models, see for example \cite{brigocvabook}. It happens however that many processes seen in finance not only have jumps (discontinuities, see for example \cite{BrigoPallaTorreWiley}) but also present the feature of long term memory (see e.g.~\cite{GGTheArt2009} for a more extensive discussion on these issues). In that case one might be tempted to address the issue of model risk when using AT1P. For this reason we have added a section on stress tests to get a feeling of what might happen with the model output in case input parameters have significant changes. Further analysis could consider the extensions of the AT1P model with random default barrier and possibly random volatility, see for example \cite{BMT2009} and \cite{brigocvabook}.

 \subsection{Calibration}

We applied the 2-step algorithm described in Section \ref{sec:calibration} considering the market data as of 15--Dec--10. 

Results are as follows: 


\begin{enumerate}
\item
The simulated annealing optimization set the output, to be used for the local optimization starting point, at

\begin{tabular}{| c| r@{.}lr@{.}lr@{.}l r@{.}lr@{.}lr@{.}lr@{.}l r@{.}l r@{.}l |}
\hline
Point &  \multicolumn{2}{|c}{B}  &  \multicolumn{2}{c}{H} &  
\multicolumn{2}{c}{$\sigma_1$}  &  \multicolumn{2}{c}{$\sigma_2$}  &  \multicolumn{2}{c}{$\sigma_3$} &  \multicolumn{2}{c}{$\sigma_4$} &
 \multicolumn{2}{c}{$\sigma_5$}  &  \multicolumn{2}{c}{$\sigma_6$} &  \multicolumn{2}{c|}{$\sigma_7$} \\
\hline
$\vet x ^{(0)}$ &   0&0002 &   0&8174 & 0&1030  &   0&0328 &  0&0007  &    0&2981 &    0&1961  &    0&2989 &   0&5006 \\
\hline
\end{tabular}•
\item
Levenberg-Marquardt set the optimum set of parameters as 

\begin{tabular}{|c| r@{.}lr@{.}lr@{.}l r@{.}lr@{.}lr@{.}lr@{.}l r@{.}l r@{.}l |}
\hline
Point &  \multicolumn{2}{|c}{B}  &  \multicolumn{2}{c}{H} &  
\multicolumn{2}{c}{$\sigma_1$}  &  \multicolumn{2}{c}{$\sigma_2$}  &  \multicolumn{2}{c}{$\sigma_3$} &  \multicolumn{2}{c}{$\sigma_4$} &
 \multicolumn{2}{c}{$\sigma_5$}  &  \multicolumn{2}{c}{$\sigma_6$} &  \multicolumn{2}{c|}{$\sigma_7$} \\
\hline
$\vet p$ &   0&0000 &   0&9533 & 0&0260  &   0&0119 &  0&0187  &    0&0187 &    0&0212  &    0&0207 &   0&0250 \\
\hline
\end{tabular}•
\end{enumerate}•

We point out that the global optimization we used is quite coarse, so that stopping our calibration at that stage would hardly be satisfactory. However, we use this global step simply to identify a region for the parameter spaces where it is good to start a more local-type optimization. We have tried several other calibration schemes and the final results we obtain are always close to the optimum parameters in the Levenberg-Marquardt algorithm above.

\subsection{Pricing application}
\label{sec:pricingCoCo}
For the pricing exercise of this paper we have used the Lloyds Bank issued Coco bond  shown in Table~\ref{tbl:features_1}.


\begin{table}
\centering
\begin{tabular}{|r | l |}
\hline
\textbf{ISIN ID:} &  XS0459088877 \\
 \textbf{Level of subordination}& Lower Tier 2 (LT2)\\
 \textbf{Issue date}& Dec -- 09\\
 \textbf{Maturity date} & Mar -- 20\\
 \textbf{Coupon}& fixed, 11.04\% per year paid twice per year, on Mar and Sept\\
\textbf{Trigger event}& Core Capital Tier 1 Ratio falling below 5\%\\
\textbf{Conversion Price}& fixed on 27th Sept 2009 at 59 GBp per share\\
\hline
\end{tabular}
\caption{Financial features of  the CoCo bond.}\label{tbl:features_1}
\end{table}

The pricing date for this example is 15--Dec--10. The instrument is actively traded in the market and on the trading date of our example had a bid-offer price per unit of nominal of  $(1.014; 1.02875)$, corresponding to a \emph{yield-to-maturity} (YTM) of $(10.52\%; 10.28\%)$. 

\begin{figure} 
\centering
\includegraphics[width = 0.8\textwidth]{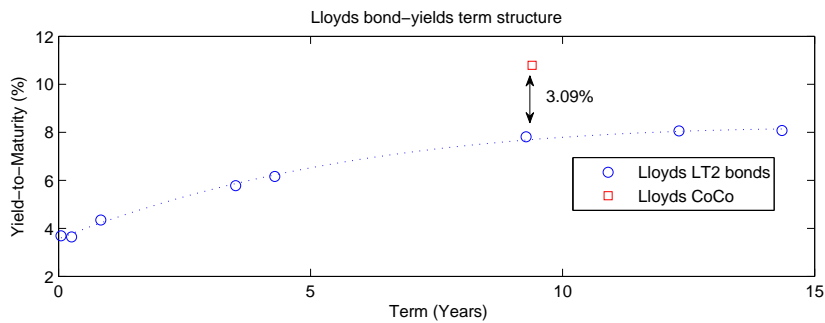}
\caption{Lloyds' bond-yields term structure as of 15-Dec-10.}
\label{fig:bondYield}
\end{figure}

Figure~\ref{fig:bondYield} shows the YTM term structure for Lower Tier 2 bonds issued by Lloyds on 15-Dec-10. Observe that the YTM of the CoCo bond is 3.09\% above the LT2 term structure. An interpretation for it is the following: 3.09\% is the additional spread that the market considers fair to pay to account for the additional risks embedded in a CoCo bond.  

\begin{table}
\centering
\begin{tabular}{| r@{/}lr@{.}l  r@{.}l@{; }r@{.}lr@{.}l |}
\hline
 \multicolumn{2}{|c}{$\Delta t$} &
    \multicolumn{2}{c}{Estimated }& 
 \multicolumn{4}{c}{Confidence } &
    \multicolumn{2}{c|}{Annualized }\\
 \multicolumn{2}{|c}{ (Years)} &
    \multicolumn{2}{c}{value}& 
 \multicolumn{4}{c}{interval $95\%$ }&
    \multicolumn{2}{c|}{YtM } \\[1ex]
\hline

  1&2 &
        {1}&{04405} & 1&03488 &  1&05321   &    {0}&{100424}\\
  1&10 &
        {1}&{03041} & 1&02196 & 1&03887 &   {0}&{102567}\\
  1&20 &
        {1}&{01775} & 1&00972 & 1&02578 &   {0}&{104589}\\ \hline
  \multicolumn{1}{c}{}& &
       \multicolumn{2}{c}{} & \multicolumn{2}{c}{} & \multicolumn{2}{c}{} &   \multicolumn{2}{c}{} \\
\hline
  1&500 &
        {1}&{02351} & 1&01599 & 1&03102 &   {0}&{103666}\\
\hline
\end{tabular}

\caption{Price, 95\%-confidence interval and annualized yield-to-maturity dependence on the sampling frequency $\Delta t$. 
%
%
}\label{tbl:price}
\end{table}

\subsubsection{Sensitivity to the sampling frequency}

In general capital ratios are supposed to be made public by companies twice per year. However, as soon as the sampling is not continuously performed, the condition~\eqref{eqn:Ptau} is not satisfied anymore. As trigger and default time can be observed only on discrete times, observing simultaneity between the two becomes more likely. Simultaneity of trigger and default time impacts on the price in two ways in the current framework:
\begin{itemize}
\item
it  increases the number of coupons received by bondholders, since default is postponed to the next discrete date;
\item
it brings to zero the value of the ``conversion recovery rate''.
\end{itemize}
With respect to the second point, this is due to the very definition of default we have in AT1P. At default $V_\tau = \hat H(\tau)$ so that the value of the barrier option through which we calculate the stock price is zero\footnote{Although the institution might review capital ratios in an annual or bi-annual basis, Rating Agency analysts and Regulators will certainly follow it much more closely. In that case either a rating agency might downgrade an institution, thus triggering a chain of events that will call the conversion trigger, or a regulatory authority might come forward and request the conversion to happen. This sort of events have not been explicitly modeled in our approach.}.

In Table~\ref{tbl:price} we show how the choice of the sampling frequency impacts the price of the CoCo bond. We fix the payout ratio $q = 0$ and the correlation parameter $\eta = 1$. We resort to Monte Carlo simulation to calculate prices. We also report the 95\%- confidence interval and the YTM. Observe that varying the sampling frequency causes a variation in price and YTM of about 2.5\% and 4\% respectively. 


The change in price that is due to the change of the sampling frequency points to the same sort of model risk that has been observed e.g. in~\cite{GGTheArt2009} (Ch.15 and the references therein) when analyzing CPDO's using the continuous Gaussian process. This suggests the necessity of additional research on the AT1P model this time using jump / Levy processes. Furthermore, a richer version of AT1P, called SBTV, with a random default barrier and possible inclusion of misreporting or fraud is introduced in \cite{BMT2009}, see also \cite{brigocvabook} and references therein.


\subsubsection{A heuristic criteria to set the sampling frequency}

{Let us consider a bond paying a fixed rate $c$ per annum on a time schedule $T^b_1,\dots, T^b_N$. Assume a null recovery rate in case of default. We can calculate this bond price per unit notional at time $t$ with $t<T^b_0 < T_1^b$ (where $T^b_0$ is the first reset date) as 
\begin{equation}\label{eqn:riskyBond}
P(t) = c \sum_{i = 1}^{N} \alpha^b_i D(t, T_i) \mathbb Q_t(\tau > T_i^b) + D(t,T^b_N) \mathbb Q_t(\tau > T^b_N), \quad \alpha^b_i := T^b_{i} - T^b_{i-1} .
\end{equation}

The AT1P model allows us to calculate analytically expressions as the one in equation \eqref{eqn:riskyBond} by using \eqref{eqn:AT1P_surv}. We can use this fact to establish a criteria to set the time step of the Monte Carlo time grid that we use to price the CoCo bond via simulation. We may request the grid to be such that the price of plain defaultable bonds obtained through the Monte Carlo engine agrees with the one calculated through the AT1P closed-form formula. Following this criteria, we found that by using $\Delta t = 1/500$ the price calculated through \eqref{eqn:riskyBond} always lies in the 95\%-confidence interval of the Monte Carlo engine. Therefore, our choice for the sampling frequency is 1/500.}

\subsubsection{Sensitivity to drift and correlation}

For the calculations in this subsection, we fixed the sampling frequency to $\Delta t = 1/500$. It is worth stressing the extreme difficulty in calibrating a crucial parameter:  the payout ratio $q$. Since a sound and robust estimation for this parameter does not appear to be immediate, we decided to run scenarios on it. The same approach has been followed for the correlation parameter $\eta$.  We calibrated our model to different values of the payout ratio $q$ and for every model we obtained in this way, we calculated the price of the CoCo bond under different correlation conditions. Results are shown in Table \ref{tbl:corrDrift}.

{  On one hand, with maximum $\eta$ and given that $\beta <0$, when $X$ grows then $c$ decreases more than with lower $\eta$. In the limit case where $\eta = 0$ and if we consider $\std (X_t)$ as constant, the capital ratio does not depend on $X_t$. In this case, it is more likely that the capital ratio does not cross the trigger ratio at all, even in scenarios where $V_t$ approaches $\hat H_t$. In those cases, however, conversion is triggered by the default  of the bond as per \eqref{eqn:Ptau}. The effect of lowering $\eta$ is an increase in the duration of the bond together with a decrease of the equity price at conversion. Table \ref{tbl:corrDrift} illustrates how, for low values of $\eta$, the price monotonically decreases as $\eta$ increases. This holds until $\eta$ reaches a level such that the trigger time is actually given by the capital ratio falling below the contractual threshold rather than the firm value crossing the barrier. In those cases one can notice that the duration is less sensitive to $\eta$, and thus the fact that the equity price at conversion is increasing in $\eta$ becomes more relevant. }


\begin{table}
\centering
\includegraphics[width = 0.7\textwidth]{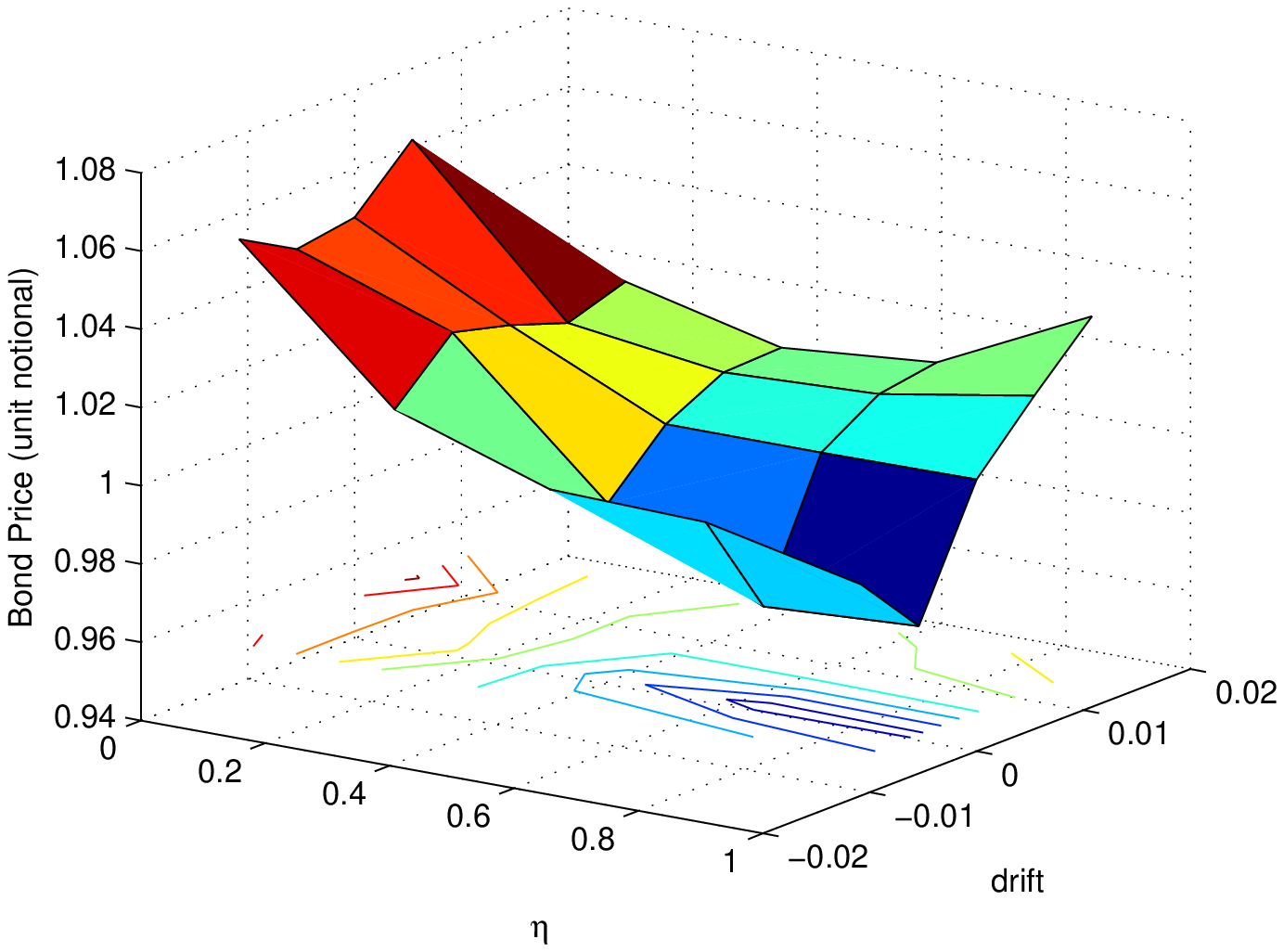}
      \bigskip \\
\begin{tabular}{| l| r@{.}lr@{.}lr@{.}l r@{.}l r@{.}l|}
\hline
 & \multicolumn{2}{|c}{$\eta = 0 $} &
    \multicolumn{2}{c}{$\eta = 0.25 $} &
    \multicolumn{2}{c}{$\eta = 0.5 $} &
    \multicolumn{2}{c}{$\eta = 0.75 $} &
    \multicolumn{2}{c|}{$\eta = 1 $}\\[1ex]
\hline

 $q = 3 r$ &{1}&{0533} &
        {1}&{0171} &
{1}&{0039} &
{1}&{0027} &
	{0}&{9940}\\
 $q = 2 r$ &{1}&{0451} &
        {1}&{0311} &
{0}&{9946} &
{0}&{9755} &
	{0}&{9777}\\
 $q = r$ &{1}&{0476} &
        {1}&{0273} &
{1}&{0092} &
{1}&{0091} &
	{1}&{0095}\\
 $q = 0$ &{1}&{0618} &
        {1}&{0221} &
{1}&{0167} &
{1}&{0184} &
	{1}&{02351}\\
 $ q =  -r$ &{1}&{0391} &
        {1}&{0271} &
{1}&{0174} &
{1}&{0208} &
	{1}&{0398}\\
\hline
\end{tabular}
\caption{Price variation given different $q$ values and different $\eta$ values. $r=0.54\% $p.a. We recall that the market traded price at the same date was $(1.014; 1.02875)$ }\label{tbl:corrDrift}
\end{table}

\subsection{Price of a Plain Defaultable Bond (PDB)}

We inferred default probabilities from the CDS market. It is indeed possible to price a plain defaultable bond using only such default probabilities and deterministic interest rates. As we have pointed out in Section \ref{sec:Calibration}, the payoff $\Pi_d(t,T)$ of the PDB with unit notional 1 can be written as:
\begin{equation}
\Pi_d(t,T) = \surv{}{T} \Pi(t,T) + \mathbbm 1_{\{\tau\leq T\}} \left( \Pi(t,\tau) +  R D(t,\tau) 1 \right), \label{eqn:plainBondPayoff}
\end{equation}
where $R$ is a deterministic recovery rate, and $\Pi(t,T)$  the discounted payoffs  of an identical risk free bond. 

We used Monte Carlo simulation to value $\expV{\Pi_d(t,T)}{t}$ and we chose a PDB with characteristics (except possibly for the conversion feature of the CoCo) that was as close as possible to the CoCo bond we valued in Section \ref{sec:pricingCoCo}. Its features are shown in Table~\ref{tbl:features_2}.

\begin{table}

\begin{tabular}{|r | l |}
\hline
\textbf{ISIN code} & XS0497187640 \\ 
\textbf{Level of subordination} & Lower Tier 2 (LT2)\\
\textbf{Issue date} & Mar -- 10\\
\textbf{Maturity date} & Mar -- 20\\
\textbf{Coupon}& fixed, 6.5\% per year paid once per year, on Mar\\
\textbf{Trigger event}& n.a.\\
\textbf{Conversion Price}& n.a.\\
\hline
\end{tabular}
\caption{Financial features of  the PDB with ISIN code XS0497187640}\label{tbl:features_2}
\end{table}
%

\begin{table}
\centering\begin{tabular}{| r@{.}lr@{.}l  r@{.}l@{; }r@{.}l |}
\hline
 \multicolumn{2}{| c}{Market Price} &
    \multicolumn{2}{c}{Estimated value }& 
 \multicolumn{4}{c| }{Confidence interval $95\%$}\\[1ex]
\hline
  0&904517 &
{0}& {922531} & 0&916056 &  0&929007  \\
\hline
\end{tabular}\caption{Price of the PDB with ISIN code XS0497187640}\label{tbl:plainBond}
\end{table}

In Table~\ref{tbl:plainBond} we show our estimation for the price of this PDB (for convenience purposes we also show the bid market price).

{ The difference between our model price and the market price of this security is an example of CDS-bond basis. There is a wide literature on this subject, and the basis itself can be measured in different ways. We refer the interested reader to \cite{Porter2009} for an extensive discussion and for a survey on credit spread measures for bonds. We may consider a correction to our CoCo bond price that takes into account the basis. If we translate the two prices in two yields to maturity, we obtain a difference of annual rates given by 27 basis points. We could use this adjustment in discount rates in order to account for the CDS/Bond basis when pricing the CoCo bond. This will be done in further work, where we will also consider explicit models for the basis.}

\subsection{Price of the CoCo Bond as a Plain Defaultable Bond}
In this section we estimate the impact of the conversion option embedded in the CoCo bond. We calculate this impact in terms of price/yield by pricing the same instrument but eliminating the conversion feature. That is, we price a plain defaultable bond (PDB) 
with the characteristics presented in Table~\ref{tbl:features_3}.   
\begin{table}
\centering
\begin{tabular}{| r | l |}
\hline
 \textbf{Level of subordination}& Lower Tier 2 (LT2)\\
 \textbf{Issue date}& Dec -- 09\\
 \textbf{Maturity date} & Mar -- 20\\
 \textbf{Coupon}& fixed, 11.04\% per year paid twice per year, on Mar and Sept\\
\textbf{Trigger event}& n.a.\\
\textbf{Conversion Price}& n.a.\\
\hline
\end{tabular}
\caption{Financial features of  the CoCo - PDB.}\label{tbl:features_3}
\end{table}

With a slight abuse of notation, we can use \eqref{eqn:plainBondPayoff} to write the discounted payoff of this instrument. Results are shown in Table \ref{tbl:coco_PDB}. As expected the inclusion of the conversion option reduces the price of the instrument. Conversely, removing the optionality from the CoCo, we get a higher price for the instrument.

\begin{table}
\centering\begin{tabular}{| r@{.}lr@{.}l@{; }  r@{.}lr@{.}l |}
\hline
    \multicolumn{2}{| c}{Estimated value }& 
 \multicolumn{4}{c}{Confidence interval $95\%$}&
 \multicolumn{2}{c| }{Annualized YtM} \\[1ex]
\hline
1& {21666} & 
1&20807 &  1&22525 &
0&076 \\
\hline
\end{tabular}\caption{Price of the CoCo - PDB, case $q=0, \eta =1$. The analogous price with conversion features in is 1.01775}\label{tbl:coco_PDB}
\end{table}

\begin{figure} 
\centering
\subfloat[Ratio between equity value at conversion ($E_\tau$) and contractual conversion price ($E^\star$). $q = 0$ and $\eta =1$.]{\includegraphics[width = 0.45\textwidth]{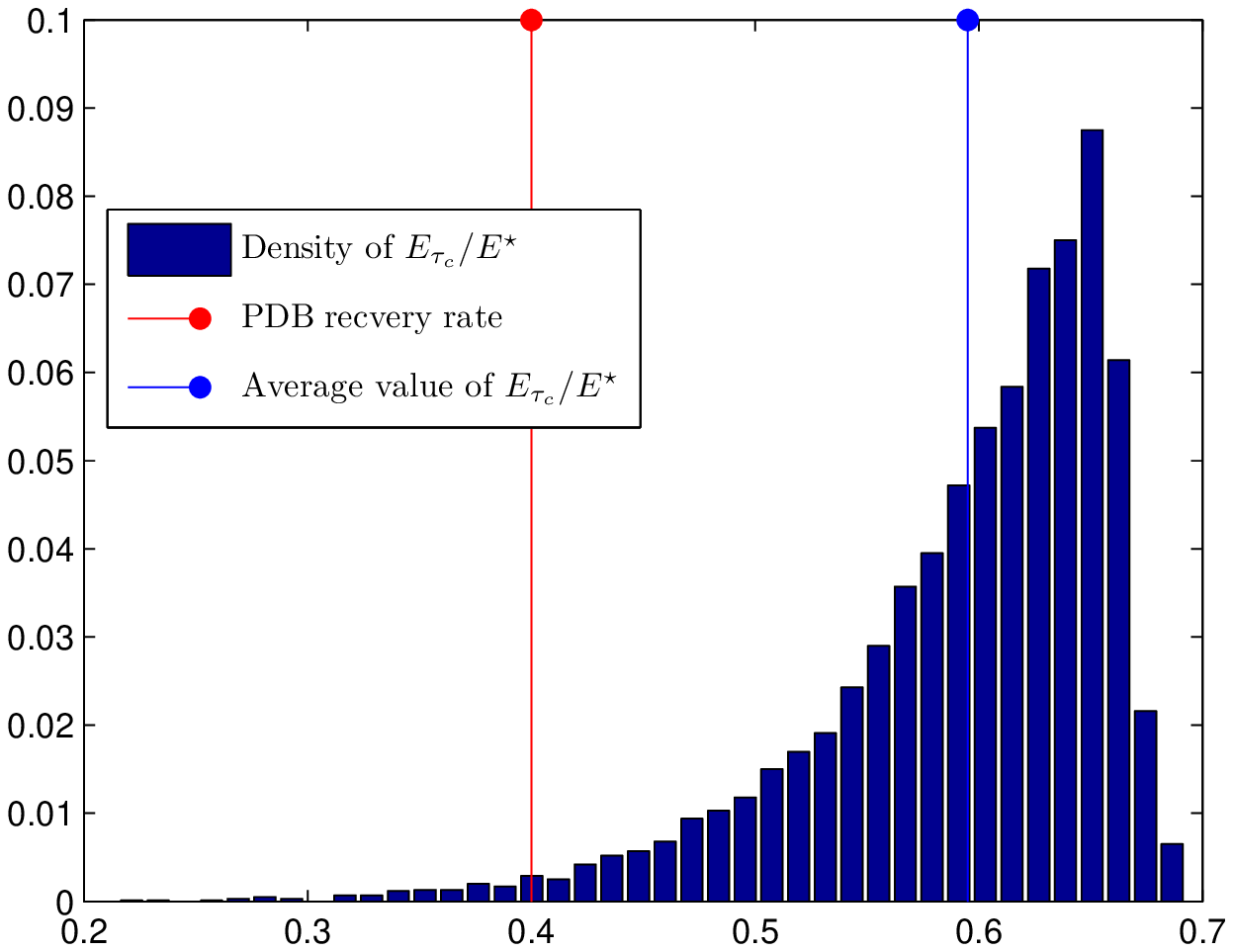}\label{fig:recRates}}
\subfloat[Conversion and default times.]{\includegraphics[width = 0.45\textwidth]{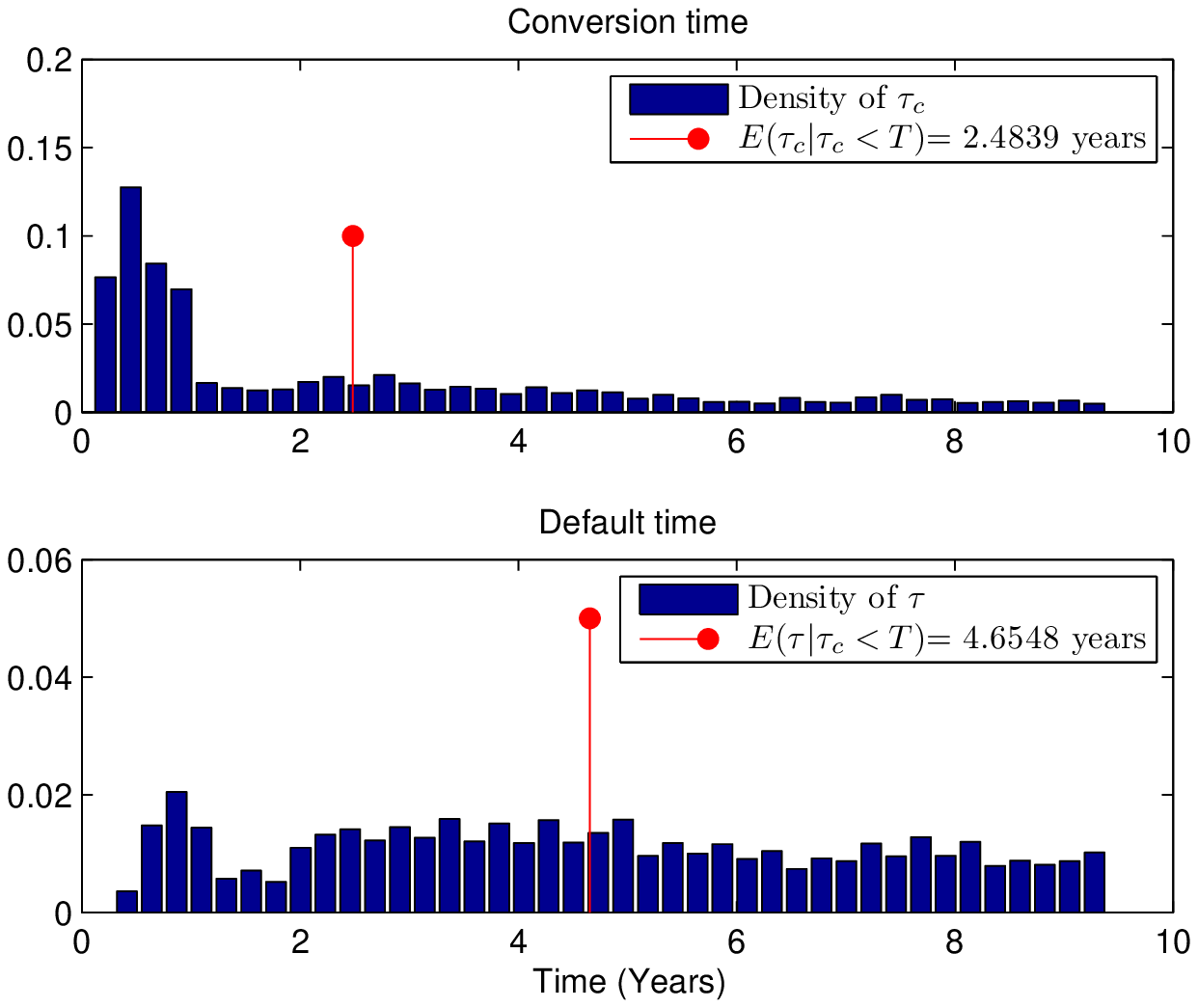}\label{fig:taus}}
\caption{Empirical distributions.}
\end{figure}

Observe that an important difference between a PDB and a Coco is what the holder receives in case of default (for the first) or conversion (for the second). The concept of \emph{recovery rate} in the case of the Coco is the ratio between equity at trigger time ($E_\tau$) and the contractually-stated conversion price ($E^\star$). In Figure~\ref{fig:recRates} we show the distribution of the mentioned ratio (that is $\frac {E_\tau}{E^\star}$) resulted of our Monte Carlo simulation. For comparison purposes we also plotted the widespread assumption of recovery rate in the CDS market (40\%). Notice that in our MC (\emph{gaussian} based) simulation the average price at conversion is 50\% of the price set in the contract. The higher recovery rate is compensated by a sensibly smaller value for the conversion time compared to the default time as shown in Figure \ref{fig:taus}.



\subsection{Stressing Input Data}
\label{subsec:StressInData}

In order to address possible issues linked to the model risk mentioned in Section~\ref{sec:pricingCoCo} we decided to keep the model as is, but to stress its parameters~\footnote{One may certainly come up with jump diffusion models. In that case however, complexity increases significantly as to our knowledge there is no closed form solutions available for calibration purpose.}. 

We have tested the price changes for variations of the CDS curve and the equity value of the company. The results are shown in Table~\ref{tbl:StressCDSAndEquity}. We moved CDS spreads up and equity prices down by 10\% and 30\% for both cases. Observe that under the model the sensitivity to equity price moves is significantly higher than to CDS spread moves. Or in another way, moves of 10\% and 30\% on equity would be equivalent to higher moves on CDS spreads. Indeed observe that a move of 10\% in equity causes a larger impact than a 30\% upper move on CDS spreads.

\begin{table} 
\centering\begin{tabular}{| r |r @{.}l r@{.}l@{; } r@{.}l |  r@{.}l r@{.}l@{; }  r@{.}l |}
\hline
  &
 \multicolumn{12}{c|}{Percentage change} \\
\cline{2-13} 
   Stressed market data  &
 \multicolumn{6}{c|}{ 10\%} &
 \multicolumn{6}{c|}{30\%} \\
 &  \multicolumn{2}{c}{ Price} &
 \multicolumn{4}{c|}{ Confidence Interval} &
\multicolumn{2}{c}{ Price} &
 \multicolumn{4}{c|}{ Confidence Interval} 
 \\[1ex]
\hline

Spot equity price &
  0&991562 &  0&983595 &  0&999528&
  0&916089 & 0&908518&  0&92366\\ 
CDS &
 1&00439 &  0&99638&  1&01241 &
0&97059 &  0&962817&  0&978362 \\

\hline
\end{tabular}\caption{Prices of the CoCo bond under stressed market data input. The percentage change has to be intended as negative when referred to the Equity spot price and positive and uniformly applied to the whole term structure when referred to CDS. The unstressed price is 1.01775}
\label{tbl:StressCDSAndEquity}
\end{table}

%

\section{Conclusions}
\label{sec:Conclusions}

In this paper we have made a detailed study of an equity / Merton based approach to price a Contingent Capital / Coco bond. The model is based on the seminal work of \cite{BrigoT2004}. As a Gaussian based model it has the advantage of analytical formulas readily available for calibration purposes. On the data side we have also used (for calibration purposes) CDS and Capital ratio estimations based on (the) proprietary database available from Fitch Solutions. Another advantage of the model is exactly the fact one does not need to have access to the whole portfolio of the institution (to have an estimation of an instrument that is in fact capital dependent).

On the technical side a possible weakness of the model would be the fact that it is Gaussian based. For addressing that issue we have come up with the sensitivity of the prices to stressed model parameters. In addition, we have considered a simple and intuitive analysis that may be helpful in highlighting poor outcomes.   
A possible continuation of this work is to adopt more realistic firm value models, possibly with jumps, and more complex models for the conversion time and its relationship with the default time. This will be considered in future work. 


\bibliographystyle{plainnat}


\bibliography{articlerefsUpdatedJ}

\begin{thebibliography}{23}
\providecommand{\natexlab}[1]{#1}
\providecommand{\url}[1]{\texttt{#1}}
\expandafter\ifx\csname urlstyle\endcsname\relax
  \providecommand{\doi}[1]{doi: #1}\else
  \providecommand{\doi}{doi: \begingroup \urlstyle{rm}\Url}\fi

\bibitem[Barucci and Del~Viva({2011})]{Barucci2011}
E.~Barucci and L.~Del~Viva.
\newblock {Dynamic Capital Structure and the Contingent Capital Option }.
\newblock \emph{{SSRN.com}}, {2011}.

\bibitem[Bielecki and Rutkowski(2002)]{BieleckiRutkowski2002}
T.~Bielecki and M.~Rutkowski.
\newblock \emph{Credit {R}isk: {M}odeling, {V}aluation and {H}edging.}
\newblock Springer Finance, Berlin, 2002.

\bibitem[Black and Cox(1976)]{BC76}
F.~Black and J.~Cox.
\newblock {Valuing Corporate Securities: some effects of bond indenture
  provisions}.
\newblock \emph{Journal of Finance}, 31:\penalty0 351--367, 1976.

\bibitem[Black and Scholes(1973)]{BS73}
F.~Black and M.~Scholes.
\newblock {The Pricing of Options and Corporate Liabilities}.
\newblock \emph{Journal of Political Economy}, 81:\penalty0 634--54, 1973.

\bibitem[Blundell-Wignall and Atkinson(2010)]{Blundell}
A.~Blundell-Wignall and P.~Atkinson.
\newblock Thinking {B}eyond {B}asel {III}: Necessary solutions for capital and
  liquidity.
\newblock \emph{OECD Journal: Financial Market Trends}, 2010.

\bibitem[Brigo(2005)]{Brigo2005}
D.~Brigo.
\newblock Market models for cds options and callable floaters.
\newblock \emph{Risk Magazine, January issue}, 2005.

\bibitem[Brigo and Alfonsi(2005)]{BrigoAlfonsi05}
D.~Brigo and A.~Alfonsi.
\newblock Credit default swap calibration and derivatives pricing with the ssrd
  stochastic intensity model.
\newblock \emph{Finance and Stochastics}, 9\penalty0 (1), 2005.

\bibitem[Brigo and Chourdakis(2009)]{BrigoChourdakis}
D.~Brigo and K.~Chourdakis.
\newblock Counterparty risk for credit default swaps: Impact of spread
  volatility and default correlation.
\newblock \emph{International Journal of Theoretical and Applied Finance}, 12
  (07):\penalty0 1007--1026, 2009.

\bibitem[Brigo and El-{B}achir(2010)]{BrigoElbachir10}
D.~Brigo and N.~El-{B}achir.
\newblock An exact formula for default swaptions' pricing in the {SSRJD}
  stochastic intensity model.
\newblock \emph{Mathematical Finance}, 20\penalty0 (3):\penalty0 365--382,
  2010.

\bibitem[Brigo and Tarenghi(2004)]{BrigoT2004}
D.~Brigo and M.~Tarenghi.
\newblock {Credit Defaulty Swap Calibration and Equity Swap Valuation under
  Counterparty Risk with a Tractable Structural Model}.
\newblock In \emph{{Proceedings of the FEA 2004 Conference at MIT}}, 2004.

\bibitem[Brigo et~al.(2010)Brigo, Pallavicini, and
  Torresetti]{BrigoPallaTorreWiley}
D.~Brigo, A.~Pallavicini, and R.~Torresetti.
\newblock \emph{Credit Models and the Crisis: A Journey into CDOs, Copulas,
  Correlations and Dynamic Models.}
\newblock Wiley, Chichester, 2010.

\bibitem[Brigo et~al.(Forthcoming, 2013)Brigo, Morini, and
  Pallavicini]{brigocvabook}
D.~Brigo, M.~Morini, and A.~Pallavicini.
\newblock \emph{Counterparty Credit Risk, Collateral and Funding with cases
  from all asset classes}.
\newblock Wiley, Chichester, Forthcoming, 2013.

\bibitem[Brigo et~al.({2010})Brigo, Morini, and Tarenghi]{BMT2009}
D.~Brigo, M.~Morini, and M.~Tarenghi.
\newblock {Credit Calibration with Structural Models and Equity Return Swap
  valuation under Counterparty Risk}.
\newblock In {Bielecki, T., Brigo, D., and Patras, F.}, editor, \emph{{Credit
  Risk Frontiers: Sub- prime crisis, Pricing and Hedging, CVA, MBS, Ratings and
  Liquidity}}, pages {457--484}. {Wiley}, {2010}.

\bibitem[De~Spiegeleer and Schoutens({2011})]{Schoutens2011}
J.~De~Spiegeleer and W.~Schoutens.
\newblock \emph{{Contingent Convertible (CoCo) Notes: Structure and Pricing }}.
\newblock {Euromoney Books}, {2011}.

\bibitem[Duffie and Singleton(1999)]{Duffie1997}
D.~Duffie and K.~Singleton.
\newblock {Modelling term structure of defaultable bonds}.
\newblock \emph{{Review of Financial Studies}}, 12:\penalty0 687--720, 1999.

\bibitem[Garcia and Goossens(2009)]{GGTheArt2009}
J.~Garcia and S.~Goossens.
\newblock \emph{{The Art of Credit Derivatives: Demystifying the Black Swan}}.
\newblock John Wiley \&\ Sons, 2009.

\bibitem[Goldstein et~al.({2001})Goldstein, Ju, and Leland]{Goldstein2001}
R.~Goldstein, N.~Ju, and H~Leland.
\newblock {An EBIT-based Model of Dynamic Capital Structure }.
\newblock \emph{{Journal of Business}}, 74\penalty0 (4), {2001}.

\bibitem[Gordy(2003)]{Gordy2003}
M.~Gordy.
\newblock {A risk-factor model foundation for ratings-based bank capital
  rules}.
\newblock \emph{J Financial Intermediation}, pages 199--232, Sept 2003.

\bibitem[Haworth et~al.({2009})Haworth, Schwarz, and Porter]{Porter2009}
H.~Haworth, N.~Schwarz, and W~Porter.
\newblock {Understanding the Negative Basis }.
\newblock \emph{{Credit-Suisse Research and analytics}}, {2009}.

\bibitem[Kothari(2006)]{Kothari06}
V.~Kothari.
\newblock \emph{{Securitization: The Financial Instrument of the Future}}.
\newblock John Wiley \&\ Sons, 2006.

\bibitem[Merton(1974)]{Merton1974}
R.~Merton.
\newblock {On the pricing of corporate debt: The risk structure of interest
  rates}.
\newblock \emph{Journal of Finance}, 29:\penalty0 449--470, 1974.

\bibitem[{Rapisarda, F.}(2003)]{Rapisarda2003}
{Rapisarda, F.}
\newblock {Pricing barriers on underlyings with time–dependent parameters}.
\newblock Technical report, available at SSRN.com, 2003.

\bibitem[Wilkens and Bethke({2012})]{Wilkens2012}
S.~Wilkens and N.~Bethke.
\newblock {Contingent Convertible ('CoCo') Bonds: A First Empirical Assessment
  of Selected Pricing Models }.
\newblock \emph{{SSRN.com}}, {2012}.

\end{thebibliography}

\end{document}